\documentclass{emulateapj}

\usepackage{graphicx}

\newcommand{\muh}{\langle \mu \rangle}

\newcommand{\Msun}{\mbox{M$_{\odot}$}}

\newcommand{\Mearth}{\mbox{M$_{\oplus}$}}

\shorttitle{Photoevaporation and the solar nebula}
\shortauthors{Mitchell and Stewart}

\begin{document}


\submitted{Received 2010 May 17; accepted 2010 August 9, Published 2010 ???}

\title{Evolution of the solar nebula and planet growth under the influence of photoevaporation}


\author{Tyler R. Mitchell and Glen R. Stewart}
\affil{Laboratory for Atmospheric and Space Physics \\
1234 Innovation Drive \\
 Boulder, CO 80303-7814, USA}


\begin{abstract}
The recent development of a new minimum mass solar nebula, under the assumption that the giant planets formed in the compact configuration of the Nice model, has shed new light on planet formation in the solar system. Desch previously found that a steady state protoplanetary disk with an outer boundary truncated by photoevaporation by an external massive star would have a steep surface density profile. In a completely novel way, we have adapted numerical methods for solving propagating phase change problems to astrophysical disks. We find that a one-dimensional time-dependent disk model that self-consistently tracks the location of the outer boundary produces shallower profiles than those predicted for a steady state disk. The resulting surface density profiles have a radial dependence of $\Sigma(r) \propto r^{-1.25^{+ 0.88}_{-0.33}}$ with a power law exponent that in some models becomes as large as $\sim \Sigma(r) \propto r^{-2.1}$. The evolutionary timescales of the model disks can be sped up or slowed down by altering the amount of far-ultraviolet flux or the viscosity parameter $\alpha$. Slowing the evolutionary timescale by decreasing the incident far ultraviolet flux, or similarly by decreasing $\alpha$, can help to grow planets more rapidly, but at the cost of decreased migration timescales. Although they similarly affect relevant timescales, changes in the far ultraviolet flux or $\alpha$ produce disks with drastically different outer radii. Despite their differences, these disks are all characterized by outward mass transport, mass loss at the outer edge, and a truncated outer boundary. The transport of mass from small to large radii can potentially prevent the rapid inward migration of Jupiter and Saturn, while at the same time supply enough mass to the outer regions of the disk for the formation of Uranus and Neptune.
\end{abstract}





\section{Introduction}

The development of the minimum mass solar nebula (MMSN) was a critical step in understanding the origin of the solar system. The MMSN was developed by \citet{weidenschilling77}, and similarly by \citet{hayashi81}, by augmenting the planets' estimated heavy element component to solar composition. The necessary mass was then distributed in annuli centered on the planets' current semimajor axis and a single power law was fit to the derived surface density constraints:
\begin{equation}
\label{eqn:mmsn}
\Sigma(r) \approx 1700\biggl(\frac{r}{1 {\rm AU}}\biggl)^{-3/2}\ {\rm g\ cm^{-2}},
\end{equation}
where $\Sigma(r)$ is the surface mass density at a radius $r$.

Since its inception, many shortcomings have been recognized in the MMSN. Observations of intermediate age ($2.5-30$ Myr) clusters indicate a mean disk lifetime of $\sim 6$ Myr; consistent with a gas dissipation timescales for circumstellar disks of $\sim 1-10$ Myr \citep{haisch01,hernandez07}. It must be emphasized that the MMSN, by definition, contains the {\it minimum} amount of mass necessary to build the planets at their current semi-major axes. Therefore, any proposed solar nebula more massive than the MMSN is allowable. Despite this, the canonical MMSN has been used for decades as the initial conditions for both disk evolution and planet formation simulations. It is difficult to grow the cores of the giant planets within the time constraint of the gas dissipation timescale given the low mass surface densities predicted for the MMSN.

The recent development of the Nice model has shed new light on the process of planet formation in the solar system \citep{gomes05,morbidelli05,tsiganis05}. The model assumes the giant planets formed in a much more compact configuration than they now reside. Simulations suggest that the giant planets migrated through scattering interactions with small planetesimals exterior to Neptune. The mass and location of this planetesimal disk play critical roles in the outcome of their simulations. These interactions caused the inward migration of Jupiter and the outward migration of the other giant planets. In their simulations, chaotic behavior in the outer solar system is initiated by the crossing of Jupiter and Saturn through their 2:1 mean motion resonance (MMR). The chaotic behavior causes the rapid outward migration of Uranus and Neptune. Uranus and Neptune switch places in about half of their simulations which nicely explains why Neptune is more massive than Uranus \citep{tsiganis05}. 

Assuming the giant planets formed in the compact configuration predicted by the Nice model, \citet{desch07} developed a new steady state disk model. The predicted disk has a much steeper profile and much larger surface densities than that predicted by the MMSN. A truncated {\it decretion} disk, characterized by outward mass transport, is required in order to maintain the steep profile in a quasi steady state. Photoevaporation was invoked as a natural mechanism for truncating the disk and removing mass at the outer boundary. It has been shown that such large surface densities can cause significant planetary migration with planets being rapidly lost into the Sun \citep{crida09}.

Star formation in giant molecular clouds implies that the majority of stars in our galaxy were born in clusters rather than in isolation. The same is most likely true for the Sun. ninety percent of stars born in clusters are born into rich clusters with 100 or more members with masses in excess of $50$ \Msun \citep{lada03}. It is estimated that the Sun formed in a cluster of $1000-10,000$ stars, which in turn implies an average external FUV flux that is a few thousand times the non-cluster background, but with a standard deviation that is comparable to the average value \citep{fatuzzo08,adams10}. The compelling reasons for the Sun being formed in a cluster of this size are (1) the abundance of the short-lived radioactive nuclide $^{60}{\rm Fe}$ derived from meteoric samples cannot be produced by spallation in the solar system and can only be explained by an extrasolar nucleosynthetic origin. The capture of this extrasolar $^{60}{\rm Fe}$ into the early solar system is more likely if the Sun's birth cluster contained a sufficient number of massive supernovae \citep{wadhwa07,adams10}. (2) Sedna's orbit requires a stellar encounter at a distance of less than or equal to $400$ AU \citep{kenyon04,morbidelli04,brasser06}. A smaller birth cluster would not yield such a close stellar encounter during its lifetime and cannot provide the necessary amount of $^{60}{\rm Fe}$ whereas a larger birth cluster would give such a large FUV flux that it would photoevaporate the disk before the giant planets can form \citep{adams10}.

Observations of low mass young stellar objects near the Trapezium cluster in Orion show disks silhouetted against the background nebula, the so called proplyds (PRoto PLanetary DiskS; Bally et al. 1998, McCullough et al. 1995). Initial modeling of these disks invoked photoevaporation as a result of ionizing radiation from an embedded central source. Observations of the same cluster show that mass loss must be due to neutral flows generated at the disks' surfaces. The outflows then become ionized at some distance from the base of the flow \citep{johnstone98}. A natural explanation for the observed neutral outflows is that the disks are heated by far-ultraviolet (FUV) radiation with energies in the range of $6-13.6$ eV. The neutral outflows then expand and are ionized by extreme ultraviolet (EUV) radiation with $E > 13.6$ eV. Models of external irradiation by nearby, massive stars were successful in explaining the observations of proplyds in Orion \citep{johnstone98}.

The mass-loss rate due to photoevaporation by an external FUV source can be readily derived using a simple, spherically symmetric model of outflow from an infinitely dense cloud of radius $r_{\rm c}$. Assuming that gas, of mean molecular weight $\muh$, is driven outward at a constant speed, $v_{\rm w}$, the mass-loss rate is
\begin{equation}
\label{eqn:approx}
\dot{M} = 4 \pi r^2 \muh v_{\rm w} n(r),
\end{equation} 
where $n(r)$ is the number density of the flow at radius $r$. The column density of the outflow, $N_{\rm H}$, is given by
\begin{equation}
\label{eqn:column_integral}
N_{\rm H} = \int^{\infty}_{r_{\rm c}} n(r) dr.
\end{equation}
Eqn. (\ref{eqn:approx}) can then be solved for $n(r)$, substituted into Eqn. (\ref{eqn:column_integral}) and integrated: 
\begin{equation}
\label{eqn:column}
N_{\rm H} = \int^{\infty}_{r_{\rm c}} \frac{\dot{M}}{4 \pi r^2 \muh v_{\rm w}} dr = \frac{\dot{M}}{4 \pi \muh v_{\rm w} r_{\rm c}}.
\end{equation}
The result is then solved for the mass-loss rate, $\dot{M}$. The mass-loss rate is dependent on the column density of attenuation, $N_{\rm H}$, and proportional to the radius, $r$.
\begin{equation}
\label{eqn:approx_loss}
\dot M = 4 \pi r_{\rm c} \muh v_{\rm w} N_{\rm H}.
\end{equation}
The outflow velocity can then be set equal to the sound speed in the FUV heated disk ``atmosphere'' which is analogous to the isothermal atmosphere in the simplified model that was derived by \citet{adams04} and will be presented in \S \ref{sec:photoevaporation}. This results in the mass-loss rate of Eqn. (\ref{eqn:massloss_b}) differing only by the geometric factor ${\mathcal F}$, which incidentally is of order unity.

A visual magnitude of extinction $A_{\rm v}$, of order unity, typically requires the column density $N({\rm H}) \approx 5\times 10^{21} {\rm cm^{-2}}$. A column density of roughly $10^{21} {\rm cm^{-2}}$ is a generally accepted value for complete extinction \citep{johnstone98,adams04}. If the disk atmosphere is heated to a temperature of 1000 K \citep{adams04}, then the gravitational radius, beyond which heated gas can escape from the Sun, is about 100 AU (see Eqn. (11)). Our estimate of the mass-loss rate is therefore $\dot{M} \approx 1.5\times 10^{-7} {\rm \Msun\ yr^{-1}}$, which is certainly enough to clear the solar nebula on the $10^6$ yr timescale needed to match observations.

External irradiation has previously been invoked to successfully explain the over abundance of noble gases in Jupiter. The Galileo spacecraft made in situ measurements of the composition of Jupiter's atmosphere. It measured an enrichment in heavy elements with respect to solar values. In particular, it measured the enrichment of the noble gases Ar, Kr, and Xe, species that only condense at temperatures of less that $100$ K. In light of these findings, it was proposed that Jupiter must have formed from low temperature planetesimals in a cold environment, well outside of Jupiter's orbit, allowing for the direct condensation of noble gases \citep{owen99}. It was then proposed that the noble gases could have been delivered by clathrates, allowing for higher formation temperatures \citep{gautier01}. Delivering a large amount of noble gases as clathrates would require a large amount of ${\rm H_2O}$ ice. Based on internal structure models of Jupiter that constrain the core mass to be $<40$ \Mearth, it is difficult to explain how sufficient amounts of noble gases were delivered to Jupiter in this fashion \citep{guillot06}.

Using the equations of mass loss from \citet{adams04}(see \S \ref{sec:photoevaporation}), \citet{guillot06} were able to explain the over abundance of noble gases in Jupiter's atmosphere as a result of the loss of hydrogen and helium through the photoevaporative process. They used a simplified one-dimensional evolutionary disk model, in which the vertical structure was averaged, to determine the enrichment of the solar nebula in heavy elements. The noble gases are assumed to be trapped within solids in the cold outer disk while hydrogen and helium are removed by photoevaporation. As solids migrated inward due to gas drag, they are then released as gases in the inner disk and incorporated into the giant planets. This scenario differs from previous models that required the noble gases to be accreted while trapped in solid planetesimals. Removing the requirement that noble gases be delivered in solids further loosens constraints on nebular temperatures in the outer solar system. \citet{guillot06} modeled photoevaporation from both EUV generated by an embedded early Sun and by ambient FUV generated by neighboring cluster members. They found that the scenario involving EUV was problematic due to the long timescale involved in the removal of nebular gas and the high constant value of EUV flux from an ``unidentified mechanism''. In the FUV scenario, they found that the observed enrichment in Jupiter's atmosphere could be matched with a wide range of values in their free parameters. They also found that for disk atmospheres heated to $T_{\rm env} > 100$ K, the disks dissipated on Myr timescales, consistent with observations.

\citet{desch07} derived a steady state {\it decretion} disk that matched surface density constraints derived by assuming the four giant planets formed in the compact configuration of the Nice model. We extend this line of investigation by modeling a time-dependent disk that suffers mass loss by photoevaporation from an external FUV source. Radiation from the central star certainly also played some role in disk evolution. Recent numerical simulations which combine photoevaporation from the central star with viscous evolution show that the majority of mass-loss is dominated by loss from the outer regions of the disk where less tightly bound material can easily escape from the system \citep{gorti09}. The mass loss rates from irradiation by the central star are generally lower than those due to external irradiation, but depending on the assumed FUV flux they can be of comparable magnitude. The FUV environment of any given star in a young cluster is highly uncertain. Due to the incomplete sampling of the stellar initial mass function, the variety of cluster sizes and the $1/r^2$ dependence of flux on the radial distance from the center of the cluster, where the most massive stars reside, the standard deviation of predicted FUV fields can be comparable to their mean values \citep{proszkow09,adams10}. When the highly variable nature of young stellar emission is also taken into account, the ratios of internal to external FUV flux become even more uncertain. Given these uncertainties, it is generally not until late times that the internal source begins to affect the inner disk; opening an inner gap and rapidly dispersing the system \citep{clarke01,gorti09}. For these reasons we have neglected UV irradiation from the Sun in this paper.

It is our goal to explore the claim that a truncated disk will produce a disk with the steep profile found by \citet{desch07}. We use the outputs from our viscously evolving disk model to investigate the growth rates for the giant planets within an evolving disk. In accordance with the Nice model, we place our growing embryos at the same orbital radii as \citet{desch07}. Once the growth rates have been determined, we then investigate Jupiter's chances of survival against migration because it is the most likely giant planet to be lost into the Sun. 

Observations of diverse extrasolar planetary systems in recent years make it evident that planetary migration plays a significant role in the planet formation process. Two types of migration have been widely investigated, commonly known as type I and type II migration \citep{ward97}. Type I migration occurs when spiral density waves are launched in a disk from an orbiting body. The density waves exert unbalanced inward and outward torques on the orbiting body. The sum of these torques is generally negative, causing a body undergoing type I migration to spiral inward on a relatively short timescale compared to the viscous timescale of the disk. Type II migration occurs when a body grows sufficiently large to open a gap in the disk. The body is then drawn along and its orbit decays on the viscous timescale of the disk. Type I migration affects smaller bodies and type II migration affects larger bodies. The cutoff between the two types of migration occurs when the Hill sphere of the orbiting body becomes comparable to the scale height of the disk. In high-mass disks, runaway migration may occur before the planet can open a gap, leading to very rapid orbital decay (type III migration). Despite the current uncertainties in migration theory, it is almost certain that it played a role in the processes that formed giant planets. Any comprehensive model of planet formation must consider the implications of migration on planet survival \citep{armitage10}.

This paper is organized in the following manner. In \S \ref{sec:desch} we will review Desch's (2007) MMSN model and it's implications for planet growth. In \S \ref{sec:photoevaporation} we will discuss the role that photoevaporation plays in the evolution of the solar nebula. In \S \ref{sec:disk} we will present our model of a time-dependent protoplanetary disk that is losing  mass due to photoevaporation by an external massive star. In \S \ref{sec:evolution}, \S \ref{sec:growth} and \S \ref{sec:migration} we present our results for evolving disks, planet growth within those disks and the chances of Jupiter's survival against migration. \S \ref{sec:discussion} discusses our results and lays out plans for future work. Finally, a brief summary will be presented in \S \ref{sec:summary}.


\section{Desch Model}
\label{sec:desch}

The known inadequacies of the MMSN model and the development of the Nice model led \citet{desch07} to re-investigate the primordial solar nebula. Within the context of the Nice model, \citet{desch07} has developed an new MMSN with Jupiter located at $5.45$ AU, Saturn at $8.18$ AU, Neptune at $11.5$ AU, and Uranus at $14.2$ AU. He assumes that Uranus and Neptune switched places during the chaotic period following the crossing of Jupiter and Saturn through their 2:1 MMR. These four mass constraints were combined with mass constraints from chondrules, the asteroid belt, and the disk of primordial planetesimals laying outside the orbit of Uranus to develop a single power-law profile for the primordial solar nebula:

\begin{equation}
\label{eqn:desch}
\Sigma(r) = 343\biggl(\frac{f_{\rm p}}{0.5}\biggr)^{-1}\biggl(\frac{r}{10 {\rm AU}}\biggr)^{-2.168} {\rm g\ cm^{-2}},
\end{equation}
where $f_{\rm p}$ is the fraction of the mass of condensible solids in planetesimals.

The surface density profile of Eqn. \ref{eqn:desch} is not only more massive but much steeper than the canonical MMSN ($\Sigma \propto r^{-3/2}$). In a steady state thin $\alpha$-disk, the surface density follows the relation $\Sigma \propto 1/\nu$. Using these assumptions, along with the $\alpha$-viscosity prescription, would imply that $\Sigma \propto T(r)^{-1} r^{-3/2}$. A surface density profile consistent with the MMSN would therefore imply a constant temperature profile throughout the disk. It is generally thought that disks are flared due to irradiation from the central star. It was shown that a flared disk with an internal radial temperature distribution of $T(r) = 150\ r_{\rm AU}^{-3/7}$ can produce a spectral energy distribution that is consistent with observations of T Tauri stars \citep{chiang97}. This has led many researchers to use a temperature profile of the form $T(r) \propto r^{-1/2}$. Again, using the same assumptions about the disk, this implies that the surface density profile should be $\Sigma \propto r^{-1}$.

\citet{desch07} first investigated whether a viscously spreading accretion disk of the type studied by \citet{lynden-bell74} would adequately match his surface density constraints as well as timing constraints on planet formation. Although the surface density constraints could be matched with a viscously spreading disk, such a disk evolves too rapidly to satisfy the constraints for planet formation. At $10$ AU, planetesimals should have formed by $0.03$ Myr, but at this time he finds densities that are an order of magnitude lower than those implied by the augmented mass of Saturn. The viscously spreading disk also has trouble matching the density profile of the new MMSN (Eqn. (\ref{eqn:desch})) at small radii for early times and at large radii for late times. \citet{desch07} concludes that the various constraints on the surface density and timing of the solar nebula are best matched with a steady state profile.

Following up on his conclusion, \citet{desch07} re-examined the equations for viscously evolving steady state disks. For a viscous disk with surface density $\Sigma$ and viscosity $\nu$., the conservation of mass yields
\begin{equation}
\label{eqn:steady_disk_mass}
\frac{\partial \Sigma}{\partial t} = \frac{1}{2 \pi r} \frac{\partial \dot{M}}{\partial r}
\end{equation}
and the conservation of angular momentum yields
\begin{equation}
\label{eqn:steady_disk_momentum}
\frac{\partial}{\partial t}(\Sigma r^2 \Omega) = \frac{1}{r}\frac{\partial}{\partial r}\biggl(\frac{\dot{M}}{2 \pi} r^2 \Omega + r^3 \Sigma \nu \frac{d \Omega}{dr} \biggr),
\end{equation}
where $\Omega$ is the angular frequency. $\dot{M}$ is the net flow of mass through an annulus of the disk with a negative mass flux corresponding to inward accretion. Integrating Eqn. (\ref{eqn:steady_disk_momentum}) results in
\begin{equation}
\label{eqn:momentum_integrated}
\frac{-\dot{M}}{2 \pi} r^2 \Omega - r^3 \Sigma \nu \frac{\partial \Omega}{\partial r} = {\rm const.}
\end{equation}
The constant  is evaluated by choosing an appropriate boundary condition. This is where \citet{desch07} diverges from previous derivations. In the past, the equation has been solved by assuming the dominant boundary is the {\it inner} boundary and the evolution of the disk is governed by the mass flux across the {\it inner} boundary while the {\it outer} boundary is allowed to expand indefinitely such that angular momentum is conserved.

Assuming a temperature profile of the form $T(r) \propto r^{-q}$, a viscosity of the form $\nu = \alpha c_{\rm s}/\Omega$, and the standard $\alpha$-viscosity prescription which will be discussed in \S \ref{sec:disk}, he finds a general solution for the surface density profile,
\begin{equation}
\label{eqn:desch_density}
\Sigma(r) = \frac{\dot{M}}{3 \pi \nu(r_0)} \biggl(\frac{r}{r_0}\biggr)^{-(3/2-q)}+\frac{\dot{m}}{3 \pi \nu(r_0)} \biggl(\frac{r}{r_0}\biggr)^{-(2-q)}.
\end{equation}

Instead of using boundary conditions to determine the constants of integration $\dot{m}$ and $r_0$, \citet{desch07} matches this solution to his derived surface density profile (Eqn. (\ref{eqn:desch})). For a steady state disk to be consistent with a surface density profile $\Sigma(r) \propto r^{-2.168}$, the mass flux $\dot{M}<0$. This implies that the solar nebula must be a {\it decretion} disk, with significant mass loss from the outer disk edge, rather than an {\it accretion} disk. Unlike a standard {\it accretion} disk which only requires sufficient outflow for the removal of angular momentum, a {\it decretion}  disk is characterized by an significant outward net flow of mass. The outward mass flow is driven by photoevaporation which requires a constant supply of mass from the inner solar system to replenish mass lost at the outer edges. Depending on local conditions, mass loss from the outer edge can dominate over accretion onto the central star necessitating an outward mass transport throughout the majority of the disk. It is also implied that the disk must be truncated at an outer edge, $r_{\rm d}$. Truncation can be naturally explained by invoking photoevaporation by an external source.

Although Desch's MMSN is an improvement over the original MMSN, it still suffers its own limitations. For one, \citet{desch07} is unable to constrain his disk's outer radius to better than within $30-100$ AU. A recent paper by \citep{crida09} also points out that the large surface densities present in Desch's model would cause substantial migration of the giant planets. Using the hydro code FARGO in its 2D1D version, their models show Jupiter quickly falling into the regime of type III runaway migration and rapidly falling into the Sun. Even though \citet{desch07} was able to construct a steady state solution, a steady state profile is an oversimplification and is inconsistent with a disk eroded by photoevaporation if the planetary accretion timescale is a significant fraction of the disk lifetime.


\section{Photoevaporation}
\label{sec:photoevaporation}

The presence of early-type stars in the vicinity of the solar nebula would have exposed it to FUV radiation. FUV radiation would have heated the periphery of the disk. Gas heated to sufficient temperatures would then have become unbound from the disk. The gravitational radius, $r_{\rm g}$, is defined as the radius at which the sound speed of the heated gas, $a_{\rm s}$, equals the escape speed from the system. Gas beyond the gravitational radius will escape from the system:
\begin{equation}
\label{eqn:radius}
r_{\rm g} = \frac{G M_* \muh}{kT}.
\end{equation}
Here $G$ is the gravitational constant, $M_*$ is the mass of the central star, $k$ is the Boltzmann constant, and $T$ is the temperature of the super-heated atmosphere, or what we will refer to as the envelope temperature, $T_{\rm env}$. The gravitational radius is the canonical radius beyond which gas heated to a temperature $T_{\rm env}$ will escape from the disk. In actuality, gas can escape from the disk at radii substantially smaller than $r_{\rm g}$.

\begin{figure}[ht]
\begin{center}
\includegraphics[width=3in]{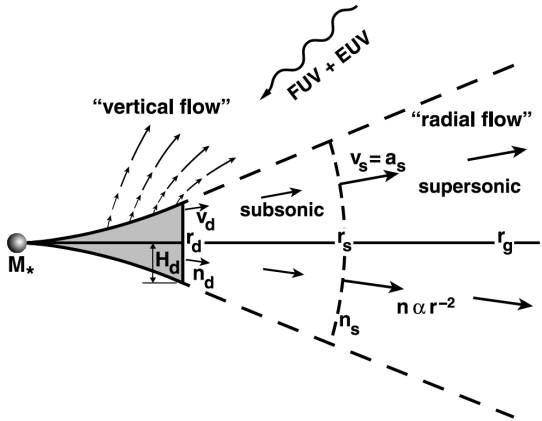}
\caption[adams]
	{
	\label{fig:adams}
(Figure taken from \citet{adams04}.) Schematic of a disk with radius $r_{\rm d}$ around a star of mass $M_*$, illuminated by the FUV (and perhaps EUV) radiation from nearby stars of grater mass. The disk is inclined so that the top and edge are exposed. The disk scale height is $H_{\rm d}$ at the outer radius $r_{\rm d}$. In the subcritical regime, where $r_{\rm d} < r_{\rm g}$, the bulk of the photoevaporation flow (the radial flow) originates from the disk edge, which marks the inner boundary. The flow begins subsonically at $r_{\rm d}$, with speed $v_{\rm d}$ and density $n_{\rm d}$. The flow accelerates to the sound speed at $r_{\rm s}$ (the sonic point), which lies inside the critical escape radius $r_{\rm g}$. Beyond the sonic point, the flow attains a terminal speed and the density falls roughly as $n \propto r^{-2}$. Although some material is lost off the top and bottom faces of the disk (the vertical flow), its contribution to the mass-loss rate is secondary to that from the edges. Nonetheless, the polar regions are not evacuated, the star is  fully enveloped by the circumstellar material, and the incoming FUV radiation will be attenuated in all directions.
	}
\end{center}
\end{figure}

In reality, the heated atmosphere has a depth-dependent temperature and the heating and resultant outflow are complicated processes. Because of these complexities, it is useful to employ a simplified model with an isothermal atmosphere. Consider a disk irradiated and heated by external FUV radiation. Depending on the strength of the FUV flux, the heated gas will reach temperatures in the range $100\ {\rm K} < T< 3000\ {\rm K}$ \citep{adams04}. As the gas heats, it expands generating a neutral outflow. The expanding outflow begins subsonically but becomes supersonic by the time it reaches the gravitational radius. This outflow is generally isotropic, but the majority of mass loss is dominated by mass loss from the outer edge of the disk. The isotropic, neutral outflow serves to shield the disk from EUV radiation that would ionize the disk and heat it to $\approx 10,000$ K. A diagram of a disk of radius $r_{\rm d}$ around a star of mass $M_*$ illuminated by an external source and the subsequent resultant outflow was presented by \citet{adams04} and is reproduced in Figure \ref{fig:adams}. This outflow is, in effect, a super-heated atmosphere that can be characterized by a single envelope temperature, $T_{\rm env}$.

Until recently, only mass loss beyond the gravitational radius has been considered. Analytic arguments and numerical experiments have shown that gas can be removed down to a radius of ($0.2\cdot r_{\rm g}$) \citep{liffman03,adams04}. Although the heated gas at these radii is prevented from directly escaping from the disk, there exists an atmosphere which can extend beyond $r_{\rm g}$. This atmosphere can be photoevaporated away and a resultant outflow will develop. The outflow will behave very much like a Parker wind. As mass is lost from the out-flowing atmosphere, it is replenished from the disk and mass is effectively lost at $r < r_{\rm g}$.  Assuming a temperature of $T_{\rm env}=1000$ K for the heated atmosphere of the solar nebula, $r_{\rm g} \sim 100\ {\rm AU}$. At $(0.2\cdot r_{\rm g}) = 20\ {\rm AU}$, the formation of planets would be effected by the photoevaporative outflow. For a more thorough discussion of subcritical mass loss, see \citet{adams04,hollenbach04}.

In an effort to calculate the probability that a solar-type star would experience sufficient photoevaporation from external irradiation that it would affect giant planet formation, \citet{adams04} investigated the mass-loss rates from circumstellar disks due to external FUV radiation. They studied the previously unexplored subcritical regime, where the outer radius of the disk, $r_{\rm d}$, is smaller than the gravitational radius. \citet{adams04} used a photodissociation region (PDR) code to determine the depth-dependent temperature of the gas based on the optical depth, density, and FUV flux. Their PDR code also included 46 chemical species and 222 chemical reactions. The chemistry is critical for determining the cooling rate of the gas. For a given radiation field strength $G_0$, disk size $r_{\rm d}$, disk temperature $T_{\rm d}$, and stellar mass $M_*$, an iterative procedure was used to determine the density at the base of the flow $n_{\rm d}$ as well as the flow speed at the inner boundary. These two quantities then determine the mass-loss rate.

In order to understand the results of their detailed numerical model, \citet{adams04} developed simple analytical models for the photoevaporative mass-loss rates for cases in which $r_{\rm d}$, the location of the outer edge of the disk, is both inside and outside the gravitational radius. These models are characterized by a single temperature $T_{\rm env}$. Although the strength of the FUV radiation field $G_0$ does not directly enter into these equations it specifies the envelope temperature which determines a unique sound speed, $a_{\rm s}$ for the isothermal atmosphere.
\begin{mathletters}
\begin{eqnarray}
& \dot{M} =  C_0 N_{\rm C} \muh a_{\rm s} r_{\rm g} \Bigl( \frac{r_{\rm g}}{r_{\rm d}} \Bigr) {\rm exp} \Bigl(- \frac{r_{\rm g}}{2r_{\rm d}}
\Bigr)  &  r_{\rm d} < r_{\rm g} \label{eqn:massloss_a} \\
& \dot{M} =  4 \pi {\mathcal F} \muh \sigma^{-1}_{\rm FUV} a_{\rm s} r_{\rm d} &  r_{\rm d} > r_{\rm g} \label{eqn:massloss_b}
\end{eqnarray}
\end{mathletters}
The first equation is for subcritical disks, and the second equation is for supercritical disks. $N_{\rm C}$ is the critical surface density of the flow and $\sigma_{\rm FUV}$ is the cross section for dust grains interacting with FUV radiation. The dust optical depth is given by $\tau_{\rm FUV} = \sigma_{\rm FUV} \cdot N_{\rm H}$. For an optical depth of order unity, $\sigma^{-1}_{\rm FUV}\approx N_{\rm H}$, where $N_{\rm H}$ is evaluated at the critical density $N_{\rm C} \sim 10^{-21}\ {\rm cm}^{-2}$. The factor $C_0$ is a constant of order unity used by Adams et al. (2004) to match their numerical and analytical solutions. It is used in our model to match the mass-loss rates for sub- and supercritical disks at a radius of $r_{\rm d}/r_{\rm g} = 0.25$. This is necessary because the mass-loss rates are sensitive functions of the strength of the FUV radiation field as well as the assumed matching point. We have matched the subcritical solution onto the supercritical solution, because the supercritical solution is better understood and well constrained. The factor ${\mathcal F}$, in the second equation, is the fraction of the solid angle subtended by the flow and is $\sim 1$ because the flow from the disk surface and edge merge at a radius between $r_{\rm d}$ and $2 r_{\rm d}$; creating a nearly spherically symmetric outflow \citep{adams04}.

We have chosen to model irradiation from an external source because of the success of \citet{johnstone98} in modeling proplyds in Orion as well as the need for a truncated disk discovered by \citet{desch07}. \citet{adams04} have shown that any incident EUV is attenuated very rapidly in the disk atmosphere at several disk radii. It photoionizes a portion of the disk atmosphere but is unable to penetrate deeply and affect the disk itself. EUV radiation can perhaps affect disk evolution at late stages when it could help to clear the gas on very short timescales. Therefore, our research has been focused on FUV radiation and as of yet has neglected the effects of EUV radiation.


\section{Disk Model}
\label{sec:disk}

Under the thin disk approximation the continuity equation for a small annulus of $\Delta r$ located at radius $r$, as $\Delta r \rightarrow 0$ for a disk with surface density $\Sigma(r,t)$ and viscosity $\nu$ is
\begin{equation}
\label{eqn:continuity}
r \frac{\partial \Sigma}{\partial t} = \frac{\partial}{\partial r}(r \Sigma v_r),
\end{equation}
where $v_r$ is the net radial velocity of material transported via viscous processes in the disk.
In the same disk, the conservation of angular momentum is described by
\begin{equation}
\label{eqn:momentum}
\frac{\partial}{\partial t}(\Sigma r^2 \Omega) + \frac{1}{r}\frac{\partial}{\partial r}(\Sigma r^3 \Omega v_r) = \frac{1}{r} \frac{\partial}{\partial r}\biggl(\nu \Sigma r^3 \frac{d \Omega}{dr} \biggr),
\end{equation}
where $\Omega$ is the Keplerian angular velocity, $\Omega = (GM/r^3)^{1/2}$.

Since its inception, the $\alpha$-viscosity prescription has been a useful tool for investigating the temporal evolution of circumstellar disks \citep{shakura73}. Using the $\alpha$-viscosity prescription, the viscosity can be parameterized in terms of the local sound speed in the disk, $c_{\rm s}$, and the local Keplerian orbital velocity, $\Omega_{\rm kep}$:
\begin{equation}
\label{eqn:alpha}
\nu = \alpha c_{\rm s} H = \alpha c_{\rm s}^2 \Omega_{\rm kep}^{-1}.
\end{equation}

Self-consistently solving for the viscosity requires solving nonlinear differential equations and is beyond the scope of the current work. The mechanisms that generate viscosity are not very well understood. Using the admittedly naive alpha prescription requires iteratively solving for the vertical and radial temperature structure. A more realistic model would also consider radiative transfer and cooling, processes regulated by poorly constrained dust opacities. 

For simplicity, we assume the viscosity is proportional to the radius of the disk such that
\begin{equation}
\label{eqn:viscosity}
\nu = \nu_0 (r/R_0),
\end{equation}
where $\nu_0$ and $R_0$ are scalings for viscosity and radius. This assumption has been used in the past by many other authors  \citep{clarke07,hartmann98b}.

The linear dependence of viscosity on radius in our model implies that the temperature profile in our disk is proportional to $r^{-1/2}$. In all of our simulations we use a temperature profile of the form $T(r) = 150\cdot r^{-1/2} $ K. This is consistent with earlier works, and in particular with that of \citet{desch07}. He uses a temperature profile from \citet{chiang97} that is of the form $T(r) = 150\cdot r^{-0.429}$ K. Evaluating the temperature profile above at $r = 1$ AU, $T_{\rm disk} =  T(1\ {\rm AU}) = 150$ K and using the alpha prescription for viscosity, we are able to determine the viscosity scaling constant.

\begin{equation}
\nu_0 = \alpha \sqrt{\frac{k T_{\rm disk}}{\muh}},
\end{equation}
where $k$ is the Boltzmann constant.

At this point, it is useful to define the variables $h$ and $g$, the specific angular momentum and torque. They are good variables to use when the viscosity is proportional to $r$ as they allow us to transform the viscous disk equation into a simple, linear differential equation with a constant coefficient \citep{hartmann98a}:
\begin{equation}
h = r^2 \Omega = (G M_{\rm p} r)^{1/2}
\end{equation}
and
\begin{equation}
g = -2 \pi r \Sigma \nu r^2 \frac{d \Omega}{d r} = 3 \pi \nu h \Sigma,
\end{equation}                                  
where $M_{\rm p}$ is the mass of the primary and $G$ is the gravitational constant. By expressing Eqn. (\ref{eqn:momentum}) in terms of $h$, it can be shown that they obey the relation
\begin{equation}
\frac{\partial g}{\partial h} = -\dot{M},
\end{equation}
where $\dot{M}$ is the outward mass flux in the disk. 

Using the above relation, the continuity equation becomes
\begin{equation}
\frac{\partial \Sigma}{\partial t} + \frac{1}{2 \pi r} \frac{\partial \dot{M}}{\partial h}
\end{equation}
which reduces to
\begin{equation}
\frac{\partial g}{\partial t} = \frac{3 \nu (G M_{\rm p} )^2}{4 h^2} \frac{\partial^2 g}{\partial h^2}.
\end{equation}

By substituting in the functional form of viscosity, Eqn. (\ref{eqn:viscosity}), the viscous disk equation can be written as
\begin{equation}
\frac{\partial g}{\partial t} = \frac{3}{4} \frac{\nu_0 G M_{\rm p}}{R_0} \frac{\partial^2 g}{\partial h^2}.
\end{equation}

We are further able to non-dimensionalize the problem resulting in a linear first order differential equation, 
\begin{equation}
\label{eqn:viscous}
\frac{\partial \tilde{g}}{\partial \tilde{t}} = \frac{\partial^2 \tilde{g}}{\partial \tilde{h}^2},
\end{equation}
where 
\begin{equation}
\label{eqn:tildeh}
\tilde{h}= \frac{h}{(G M_{\rm p} R_0)^{1/2}},
\end{equation}

\begin{equation}
\label{eqn:tildeg}
\tilde{g}= \frac{g}{3 \sqrt{2 \pi} N_{\rm c} \nu_0 \sqrt{G M_{\rm p} R_0} \muh},
\end{equation}
and
\begin{equation}
\label{tildet}
\tilde{t} = \frac{4 \nu_0 t}{3 R_0^2}
\end{equation}
are the non-dimensional variables.

As with most previous disk models, we employ a zero torque inner boundary condition. This allows for accretion from the disk onto the Sun. The outflowing material carries some finite amount of angular momentum away from the system and must therefore exert a torque on the outer edge of the disk. The torque exerted on the outer edge of the disk can be expressed as 
\begin{equation}
{g}_{\rm d} = 3\sqrt{2\pi}N_{\rm c}\muh\nu_0\biggl(\frac{r_{\rm d}}{R_0}\biggr)h_{\rm d}.
\end{equation}
We use this torque as the outer boundary condition in our models. In terms of our non-dimensionalized variables, the outer boundary condition is 
\begin{equation}
\tilde{g}_{\rm d} = \tilde{h}_{\rm d}^3.
\end{equation}

It must be noted that these two boundary conditions are used to solve the equation governing the temporal evolution of the mass surface density. We have, in effect, a second outer boundary condition that governs the temporal evolution of the location of the outer boundary. This condition is set by a differential equation that was derived using mass conservation at the outer boundary.

\begin{equation}
\label{eqn:boundary_mass}
\dot M_{\rm boundary \atop motion} = \dot M_{\rm viscous \atop spreading} - \dot M_{\rm photo- \atop evaporation}
\end{equation}

The mass flux due to viscous processes is simply $- \partial g/\partial h$ and the mass flux due to photoevaporation is taken to be Eqn. (\ref{eqn:massloss_a}) or Eqn. (\ref{eqn:massloss_b}) depending on the location of $r_{\rm d}$. The mass flux due to the motion of the boundary is $2 \pi r_{\rm d} \Sigma_{\rm d} (dr_{\rm d}/dt)$. The equations governing the location of the outer boundary was derived by substituting these expressions into Eqn. (\ref{eqn:boundary_mass}). Depending on the location of the outer boundary, we will either be in the subcritical regime,
\begin{equation}
\label{eqn:subcritical}
\frac{d r_{\rm d}^2}{d t} = - \frac{1}{\sqrt{2 \pi} N_{\rm c} \muh}\frac{\partial g}{\partial h_{\rm d}} - \frac{C_0 a_{\rm s} r_{\rm g}}{\sqrt{2 \pi}} \biggl( \frac{r_{\rm g}}{r_{\rm d}} \biggr) {\rm exp}\biggl[-\frac{r_{\rm g}}{r_{\rm d}}\biggr] 
\end{equation}

or the supercritical regime,

\begin{equation}
\label{eqn:supercritical}
\frac{d r_{\rm d}^2}{d t} = -\frac{1}{\sqrt{2 \pi} N_{\rm c} \muh} \frac{\partial g}{\partial h_{\rm d}} - \sqrt{2 \pi}\ a_{\rm s} r_{\rm d}.
\end{equation}

Expressed in terms of our non-dimensional variables, the motion of the outer boundary in the subcritical regime is
\begin{equation}
\label{eqn:sub_nondim}
\frac{d \tilde{h}_{\rm d}}{d \tilde{t}} = -\frac{1}{\tilde{h}_{\rm d}^3} \biggl(\frac{r_{\rm g}}{R_0}\biggr)^{1/2} \frac{\partial \tilde{g}}{\partial \tilde{h}_{\rm d}} - \frac{C_0 a_{\rm s} r_{\rm g}}{3 \sqrt{8 \pi}\ \nu_0\ \tilde{h}_{\rm d}^5} \biggl(\frac{r_{\rm g}}{R_0}\biggr) {\rm exp}\biggl[-\frac{r_{\rm g}}{2 R_0 \tilde{h}_{\rm d}^2} \biggr],
\end{equation}
and in the supercritical regime is
\begin{equation}
\label{eqn:super_nondim}
\frac{d \tilde{h}_{\rm d}}{d \tilde{t}} = -\frac{1}{\tilde{h}_{\rm d}^3} \biggl( \frac{r_{\rm g}}{R_0} \biggr)^{1/2} \frac{\partial \tilde{g}}{\partial \tilde{h}_{\rm d}} - \frac{\sqrt{8 \pi}\ a_{\rm s} R_0}{3\ \nu_0\ \tilde{h}_{\rm d}}.
\end{equation}

Using the formalism of the variable space grid (VSG) method of \citet{kutluay97} (see the Appendix for details), Eqn. (\ref{eqn:viscous}) can be discretized with an explicit finite difference method:
\begin{equation}
\label{eqn:viscous_discretized}
\tilde{g}_i^{m+1} = \tilde{g}_i^{m} + \Biggl(\frac{\tilde{h}_i^m\ \dot{s}_m\ \Delta \tilde{t}}{2\ s_m\ \Delta \tilde{h}} \Biggr)(\tilde{g}_{i+1}^m - \tilde{g}_{i-1}^m) + r(\tilde{g}_{i+1}^m - 2\tilde{g}_{i}^m + \tilde{g}_{i-1}^m)
\end{equation}
where $r$ is defined as $\Delta \tilde{t}/(\Delta \tilde{h})^2$.
The equations governing the location of the boundary (Eqn.'s (\ref{eqn:sub_nondim}) and (\ref{eqn:super_nondim})) can then also be discretized with the same explicit finite difference scheme,
\begin{equation}
\label{eqn:sub_discretized}
s_{m+1} = - \frac{\Delta \tilde{t}}{2\ \Delta \tilde{h}\ s_m^3}\biggl(\frac{r_{\rm g}}{R_0}\biggr)^{1/2}(3\tilde{g}_N^m - 4\tilde{g}_{N-1}^m + \tilde{g}_{N-2}^m) - \Delta \tilde{t}\frac{2\ C_0 a_{\rm s} r_{\rm g}}{3 \sqrt{8 \pi}\ \nu_0} \frac{r_{\rm g}}{R_0} \frac{1}{s_m^5} {\rm exp}\biggl[-\frac{r_{\rm g}}{R_0} \frac{1}{s^2}\biggr] + s_{m}
\end{equation}
and
\begin{equation}
\label{eqn:super_discretized}
s_{m+1} = - \frac{\Delta \tilde{t}}{2\ \Delta \tilde{h}\ s_m^3}\biggl(\frac{r_{\rm g}}{R_0}\biggr)^{1/2}(3\tilde{g}_N^m - 4\tilde{g}_{N-1}^m + \tilde{g}_{N-2}^m) - \Delta \tilde{t}\frac{\sqrt{8 \pi}\ a_{\rm s} R_o}{3 \nu_0\ s_m} + s_m.
\end{equation}

At each time step, depending on whether we are in the sub- or supercritical regime, either Eqn. (\ref{eqn:sub_discretized}) or Eqn. (\ref{eqn:super_discretized}) must first be solved to determine the new boundary location. Then the viscous evolution, Eqn. (\ref{eqn:viscous_discretized}) is solved. All of the simulations presented here were performed on a Mac G5 PowerPC. They were evolved on a grid of 200 points evenly spaced in specific angular momentum. 

Our initial conditions are that of the similarity solutions of \citet{lynden-bell74}. 
\begin{equation}
\label{eqn:initial}
\Sigma_{\rm init} = \frac{M_0}{2 \pi R_1 r}{\rm exp}\biggr[-\frac{r}{R_1}\biggr]
\end{equation}
where $R_1$ is the initial disk scaling radius and $M_0$ is the initial disk mass.


\section{Disk Evolution}
\label{sec:evolution}

The two free parameters of our model that most directly influence the timescales of disk evolution and planet formation are the temperature of the super-heated atmosphere, $T_{\rm env}$, and the non-dimensional viscosity parameter $\alpha$. These are also the two parameters that are the least constrained. We have performed a number of simulations to investigate the role played by these two parameters in affecting disk morphology and evolution timescales. The various input parameters and resultant timescales are tabulated in Table \ref{tab:runs}. Our reference model best matches the parameters, temperature, viscosity, etc., of Desch's model and canonical disk models in general. The reference model has an envelope temperature of $T_{\rm env} = 600\ {\rm K}$ and a viscosity parameter of $\alpha = 0.001$. 

\begin{table}[ht]
\caption[slopes]
	{
	\label{tab:runs}
	\label{secondtab}
Input parameters for various models.
}
\begin{center}
\item[]\begin{tabular}{cccc} 
\tableline 
\tableline
    simulation  & $T_{\rm env}$ [K] & $\alpha$ & normalized timescale \\
\tableline
    LV		& $600$ 	& $0.0001$	& $3.6$ \\   
    LT		& $100$		& $0.001$	& $3.6$ \\   
    Reference 	& $600$		& $0.001$	& $1.0$  \\   
    HT		& $3000$	& $0.001$	& $0.45$ \\   
    HV		& $600$		& $0.01$	& $0.81$ \\   
\tableline
\end{tabular}
\end{center}
\end{table}

In addition to our reference model, a number of simulations have been completed in order to investigate the effect that the adjustment of key parameters can have on the evolution of the disk and on planet formation timescales within that disk. For these simulations, we have varied the envelope temperature and viscosity parameter $\alpha$ to their extreme values as predicted by canonical disk and complex PDR models. 

The viscosity parameter $\alpha$ in protoplanetary disks is considered to lie within an order of magnitude of $0.001$. We have therefore considered values of $\alpha$ between $0.01$ and $0.0001$. According to complex PDR models that have been used to determine temperature and density profiles of photoevaporating outflows, the envelope temperature lies in the range $100-3000\ {\rm K}$ \citep{adams04}. The given range of temperatures corresponds to FUV fields with $300 < G_0 < 3000$ \citep{adams04}. $G_0$ is a dimensionless quantity expressed in terms of the Habing field ($1\ {\rm Habing\ field} = 1.2 \times 10^{-4}\ {\rm erg}\ {\rm cm}^{-2}\ {\rm s}^{-1}\ {\rm sr}^{-1}$) and where $G_0 = 1.7\ {\rm Habing\ fields}$ for the local interstellar FUV field \citep{tielens05}. In an effort to constrain the behavior of our photoevaporative disks, we have used only extremes in these values. The models have been labeled according to which parameter(s) have been varied, with L or H referring to either low or high values of either the viscosity (V) or the temperature (T). 

From these simulations, we have determined the time evolution of the surface density at the locations of the four giant planets in the compact configuration of the Nice model. Using the time-dependent surface densities, we then estimate the growth rates of the giant planets' cores. The growth rates and decaying surface density profiles were then used to calculate the migration rates of the giant planets.

In general, the evolution of the disks produced by our numerical simulations begin with a prolonged phase where the outer boundary expands as viscosity transports mass outward from the massive inner disk. As the disk evolution continues, it is characterized by the slow erosion of the outer boundary and a nearly self similar shape of the disk's radial mass surface density profile. One feature all disks have in common is a mass front located at a truncated outer boundary.

With the exception of the reference model, the models were run until the mass, $M_{\rm comp}$, of the gas disk within the region of giant planet formation, $2\ {\rm AU} < r < 30\ {\rm AU}$, had reached a given value. We have chosen this region because it allows us to compare our models to that of \citet{desch07}. His steady state disk model contains $\sim 0.07\ \Msun$ in the comparison region. Our models were run until $M_{\rm comp} = 0.07\ \Msun, 0.035\ \Msun\ {\rm and}\ 0.0175\ \Msun$. The reference model was additionally run until $M_{\rm comp}=0.00875$ \Msun. It took the our reference model $0.70$ Myr to evolve from a mass of $0.07\ \Msun$ to a mass of $0.035\ \Msun$ and $0.38$ Myr to evolve from there to a disk mass of $0.0175\ \Msun$. Throughout this work, all times listed for a given specified model are relative to the time when that particular model contains $M_{\rm comp} = 0.07\ \Msun$. 

\begin{figure}[ht]
\begin{center}
\includegraphics[width=2.5in]{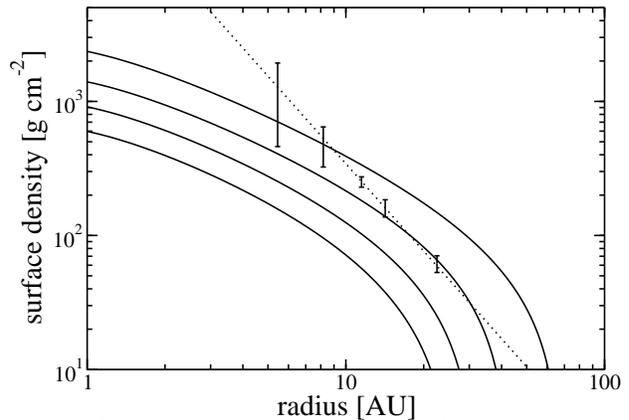}
\caption[sigma_R1]
	{
	\label{fig:sigma_R1}
The radial mass surface density at four times for the reference model. Outputs for all simulations look very similar these profiles due to our constraint on the mass contained within the planet forming region. These surface densities evolved over a $1.3$ Myr time interval. The times plotted, relative to our first output, are as follows; $0.70$ Myr, $1.1$ Myr, and $1.3$ Myr. The surface density constraints inferred from the Nice model are over plotted \citep{desch07}. The dotted line is the surface density derived by \citet{desch07}.
	}
\end{center}
\end{figure}

Snapshots of the evolving mass surface density of the reference model are shown in Figure \ref{fig:sigma_R1} at four different times that span a $1.3$ Myr interval. The surface density constraints from \citet{desch07}, as well as his derived surface density profile, have been overplotted for comparison. The uncertainty in the solid component of the two inner giant planets, with Jupiter in particular, is large because of the inability of hydrostatic models, which rely on high-pressure equations of state, to constrain current core masses. By inspection, one can see that as the disk evolves it matches with the surface density constraints of \citet{desch07} at various radii at various times. The inner surface density constraints are matched early on and the outer constraints are matched at later times. The ad hoc profile of Eqn. (\ref{eqn:desch}) is plotted with the dotted line. It is interesting to note that although our first output profile contains the same amount of mass as his profile between $2$ and $30$ AU, his profile is much steeper than our model profiles. Our surface density profiles have $\Sigma(r) \propto r^{-1.25^{+ 0.88}_{-0.33}}$, where the range of exponent here is not the uncertainty but represents rather a range of profile slopes. All profiles were fit with a power-law slope through the giant planet-forming region, $5\ {\rm AU}< r < 15\ {\rm AU}$.

Generally speaking, we have two families of disks, those with contracted radial surface density profiles and those with extended profiles. The contracted disks are dominated by photoevaporation and have slopes with an average power-law slope of $-1.6$, whereas the extended profiles are dominated by viscous spreading and have an average slope of $-0.94$. The models with contracted profiles and those with extended profiles both have slopes that slightly increase with time as the outer disk radius moves inward. The steepest profile, and our most contracted ($r_{\rm d} \approx 19$ AU), is from the simulation LV and has a radial surface density profile of $r^{-2.1}$, which is as steep as that derived by \citet{desch07}. Although the slope of the radial surface density profile of simulation LV matches that derived by \citet{desch07}, it matches at only very late times and does not maintain a steep slope such as implied by Desch's quasi steady state model. The differences between most of the derived surface density profiles of this work and of \citet{desch07} probably arise from the different assumptions that were made in these models. This will be discussed further in \S \ref{sec:discussion}.

\begin{figure}[ht]
\begin{center}
\includegraphics[width=2.5in]{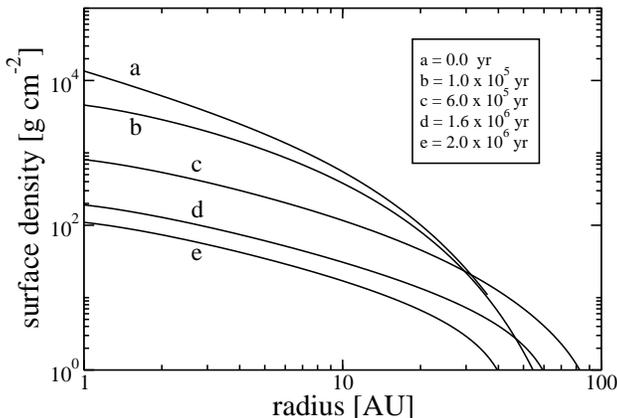}
\caption[small_disk]
	{
	\label{fig:small_disk}
The radial mass surface density at five times for the a model with an initial disk mass of $0.1\ \Msun$. These surface densities evolved over a $2.0$ Myr time interval. The times plotted, relative to our first output, are shown in the inset box. This model also uses the similarity solutions of \citet{lynden-bell74} as the initial conditions.
	}
\end{center}
\end{figure}

The initial mass is different for each of our models. This was necessary such that each model exhibited the same behavior at the times of interest. This has however required us to use some disks with unrealistically large masses, some as large as a $0.5\ \Msun$. Such large disks would likely be susceptible to gravitational instabilities. We have performed an additional simulation with a small initial disk mass ($M_0 = 0.1\ \Msun$) so that the behavior exhibited by the high-mass models can be verified under more physically realistic conditions. A plot of the radial surface density profiles at various times (times shown in inset) is shown in Figure \ref{fig:small_disk}. The elapsed time between the first and last outputs is $2.0$ Myr. As with all of our other models, this model uses the similarity solutions of \citet{lynden-bell74} as the initial conditions. In this case, the initial disk mass is $0.1$ \Msun and the scaling radius, $R_1$, has been set to $10$ AU. 

Again, as with our other models, the disk begins with a rapid expansion of the outer boundary. As the disk spreads the slope of its radial surface density profile quickly approaches a power law with an exponent $\approx -1.05$. This is similar to the average slope from all other models presented here. At $t \sim 6.6 \times 10^5$ yr the outer boundary of the disk reaches its maximum value then begins the shrink. The disk then shrinks, both in radius and in overall magnitude, as the outer edge is eaten away by photoevaporation. The disk maintains a nearly self-similar shape until the outer boundary shrinks considerably at which point the slope of the radial surface density profile steepens slightly. At the end of this simulation the disk contains $0.0035$ \Msun. We infer from this model that the behavior seen in our models is independent of disk mass.

Simulations LT and HT were designed to test the effect of FUV radiation on the evolution of the disk and in turn on planetary formation timescales. There are few, if any, strong constraints on the FUV environment of the early solar system. We therefore explored a range of disk envelope temperatures as the envelope temperature certainly affects the timescale for disk evolution. It was expected that a larger FUV flux and hence a larger envelope temperature would cause the disk to evolve more quickly. Furthermore, it was believed that a lower FUV flux and envelope temperature would result in a prolonged disk evolution allowing more time for the outermost giant planet cores to form. 

Compared to the reference model, the evolutionary timescale was in fact larger in LT and smaller in HT. It took the disk in LT $3.6$ times as long to evolve from a mass of $0.07\ \Msun$ to a mass of $0.0175\ \Msun$ as the reference model disk, whereas the disk evolution in HT was faster than that in the reference model by a factor of $0.45$.

Given the wide range of $\alpha$ generally used in solar nebula models, we varied the value of $\alpha$ to see how it affects planetary growth timescales. The viscosity parameter $\alpha$ was varied to a value of $0.0001$ for LV and to a value of $0.01$ for HV. As expected, the simulations evolved more rapidly for increasing values of $\alpha$. The evolutionary timescale of LV was $3.6$ times longer than in our reference model and the timescale of HV was $0.81$ times shorter than that of the reference model. The viscosity parameter has been changed by an order of magnitude and shows that, at the reference temperature, the disk's temporal evolution is just as  dependent on the viscosity as the temperature.

\begin{figure}[ht]
\begin{center}
\includegraphics[width=2.5in]{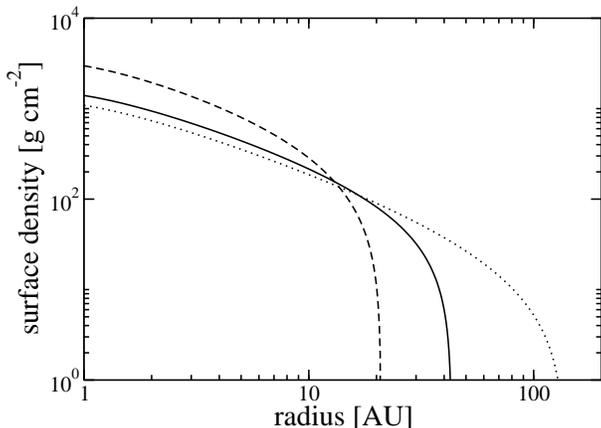}
\caption[profiles]
	{
	\label{fig:profiles}
 	Surface density profiles of the extended disk for LT, LV and the reference model. The reference model is shown in solid, LT with a dotted line and LV with a dashed line. Each disk radial surface density profile corresponds to a time when the mass contained within the planet forming region, $2\ {\rm AU} < r <30\ {\rm AU}$, is $0.035$ \Msun. These correspond to  times of $2.0$ Myr, $2.1$ Myr, and $0.70$ Myr for LV, LT, and the reference model respectively.
	}
\end{center}
\end{figure}

It is interesting to note that while LV and LT both produce longer evolutionary timescales than the reference model, the surface density profiles generated by them are strikingly different. The low viscosity of model LV prevents the mass in the inner regions of the disk from spreading outward and maintains the outer edge of the disk within a few times $10$ AU. In contrast, the relatively high reference viscosity ($\alpha=0.001$) of LT allows for the massive inner disk to rapidly spread outward to $>100\ {\rm AU}$ where it is slowly eroded by the photoevaporative outflow. The radial profiles of LV and LT can be seen in Figure \ref{fig:profiles}, along with the reference model for comparison. Each disk radial surface density profile corresponds to a time when the mass contained within the planet forming region, $2\ {\rm AU} < r <30\ {\rm AU}$, is $0.035$ \Msun. These correspond to  times of $2.0$ Myr, $2.1$ Myr, and $0.70$ Myr for LV, LT, and the reference model respectively. Here one can clearly see the difference in the radial distribution between these two models. A similar dichotomy is seen with simulations HV and HT. They both produce short evolutionary timescales relative to our reference model, but are opposite with regards to the radial extent of the surface density profiles they produce.

\section{Embryo Growth}
\label{sec:growth}

If one assumes that planets form via the accumulation of smaller bodies and not through direct gravitational collapse, the early stage of planetesimal accretion is characterized by a period of runaway growth \citep{wetherill89}. During runaway growth the velocity distribution of planetesimals is dominated by interactions with other planetesimals. During this time the velocity dispersion of planetesimals is low and gravitational focusing is effective. While gravitational focusing is effective, the largest bodies grow much more rapidly than smaller bodies and a bimodal size distribution is achieved. 

This is followed by a phase of oligarchic growth, where the velocity distributions are dominated by interactions of the larger planetary embryos. During oligarchic growth the presence of large bodies enhances the velocity dispersions of smaller bodies and decreases the velocity dispersion of the largest bodies. The increased dispersion in velocities of the smaller planetesimals decreases the effect of gravitational focusing and the largest bodies begin to decrease their growth rate. The system becomes dominated by a few large bodies, an oligarchy, separated by a few mutual Hill radii. Since our simulations take place while large amounts of gas are present, we only consider runaway growth.

Early analytical models, by \citet{safronov69} and others, overestimated the growth timescale of planets by upwards of 5 orders of magnitude and were inconsistent with observational constraints of protoplanetary systems that showed the removal of gas in $\sim 5-10$ Myr. \citet{lissauer87} developed an analytic model for the runaway growth of planetary embryos (Eqn. 3 from \citet{lissauer87}). To evaluate the growth rates of the giant planet embryos, we use Eqn. (14) of \citet{lissauer93}. The growth rate $\frac{dM_{\rm e}}{dt}$ is defined for an embryo of mass $M_{\rm e}$ and radius $R_{\rm e}$ and escape velocity $v_{\rm esc}$ embedded in a swarm of planetesimals with a local surface density $\Sigma_{\rm p}$ and velocity dispersion $\sigma$:
\begin{equation}
\label{eqn:growth}
\frac{dM_{\rm e}}{dt} = \frac{\sqrt{3}}{2} \Sigma_{\rm p}(t)\ \Omega_{\rm kep} \pi R_{\rm e}^2 \biggl(1+\frac{v_{\rm esc}^2}{\sigma^2}\biggr).
\end{equation}
The numerical prefactor depends on the velocity distribution of planetesimals and many values have been quoted in the literature, the value used here of $\sqrt{3}/2$ is due to an isotropic velocity distribution. We make the conservative assumption that the surface density of solids, $\Sigma_{\rm p}$, is $0.014$ times the surface mass density of gas and that the solids-to-gas ratio does not change with time or radius. 

One can see from Eqn. \ref{eqn:growth} that the growth rate is dependent on the geometric radius of the embryo, $\pi R_{\rm e}^2$, enhanced by a gravitational focusing factor, $(1+\frac{v_{\rm esc}^2}{\sigma^2})$. The exact value of the gravitational focusing factor has been the subject of much study over the years and is still much debated. It has been studied with both analytical and numerical studies in a variety of different regimes including gas-free and gas-damped accretion.

Numerical experiments show that the eccentricity and inclination of the planetesimals in a swarm are damped due to the interactions with gas in a disk \citep{kokubo96,kokubo98,kokubo00,kokubo02}. The damping of inclination and eccentricity due to gas drag causes, at least at the small end of the size distribution, the planetesimals to be in the shear-dominated regime where gravitational focusing is important \citep{rafikov04}. We can define a characteristic velocity
\begin{equation}
\label{eqn:hillvelocity}
v_{\rm H} = \sqrt{\frac{G M_{\rm e}}{r_{\rm H}}},
\end{equation}
based on the definition of the Hill radius,
\begin{equation}
\label{eqn:hillradius}
r_{\rm H} = a \Biggl(\frac{M_{\rm e}}{3 \Msun}\Biggr)^{1/3},
\end{equation}
where $a$ is the semi-major axis of the embryo. Our characteristic velocity marks a transition between the shear-dominated and dispersion-dominated regimes. When the velocity dispersion of planetesimals is smaller than our characteristic velocity, $\sigma < v_{\rm H}$ accretion will proceed in the shear-dominated regime \citep{armitage10}.  We adopt the characteristic velocity, $v_{\rm H}$, for the value of the velocity dispersion of planetesimals, $\sigma$, in all of our calculations of embryo growth. When $\sigma < v_{\rm H}$, the system is in the shear dominated regime and three-body dynamics become important. Therefore, Eqn. (\ref{eqn:growth}) is not strictly valid as it is derived considering only two-body effects. Also, as the embryos grow the system will transition to a dispersion-dominated regime where the embryos will grow in an oligarchic fashion. Due to the uncertainties in when this transition occurs we have focused only on runaway growth. Owing to the large variation in estimates planetary formation timescales and the wide array of unknown parameters, disk mass, viscosity, gas/solid ratio, etc., our calculations are not meant to definitively describe embryo growth but to illustrate how various surface density profiles determined by viscous evolution and photoevaporation effect planet growth. In this regard, the following results on embryo growth should be treated with some caution.

In order to determine the cores' growth rates, it was necessary to determine the time evolution of the mass surface density. The temporal evolution of surface densities at the location of each core were fit with decaying exponentials. This seemed to give a good fit to the data. These fits were then used in Eqn. (\ref{eqn:growth}) to determine the masses of the giant planet embryos as functions of time. It should be noted that in our models it is the time-dependent surface density that determines the growth rates and hence the embryo masses. In our models, the embryos initially grow rapidly because the surface densities are large, but the growth rate then begins to wane because the disk evolves and the surface densities become small. This differs from most models in which the growth rates are large throughout the duration of the simulations because they assume unrealistically large, steady state surface densities. Our research indicates that, because of the similarity of relevant timescales, planet formation models must take into account the time-dependent behavior of the solar nebula.

\begin{figure}[ht]
\begin{center}
\includegraphics[width=3in]{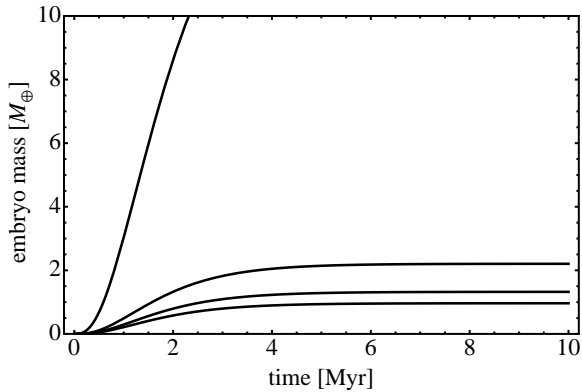}
\caption[growth run1]
	{
	\label{fig:growth1}
	Growth of giant planet embryos in our reference model. The masses of planet embryos, plotted from top to bottom, are Jupiter, Saturn, Neptune and Uranus.
	}
\end{center}
\end{figure}

Embryo masses for the four giant planets as a functions of time, determined for our reference model, are shown in Figure \ref{fig:growth1}. We were able to quite easily build the core of Jupiter to $>10$ \Mearth well within the $\sim 5-10$ Myr window implied by gas dissipation. One can see from Figure \ref{fig:growth1} that, in our reference model, we are unable to grow the cores of Saturn, Neptune and Uranus to $10\ \Mearth$ within the $\sim 5-10$ Myr time constraint. The gas simply dissipated too quickly for them to form in the allotted time. \citet{desch07} was able to grow cores of sufficient mass in his models because he relied on steady state models with surface mass densities that remained large throughout the planetary growth process. We feel that our decaying, time-dependent model is a more realistic representation of the solar nebula. 

\begin{figure}[ht]
\begin{center}
\includegraphics[width=3in]{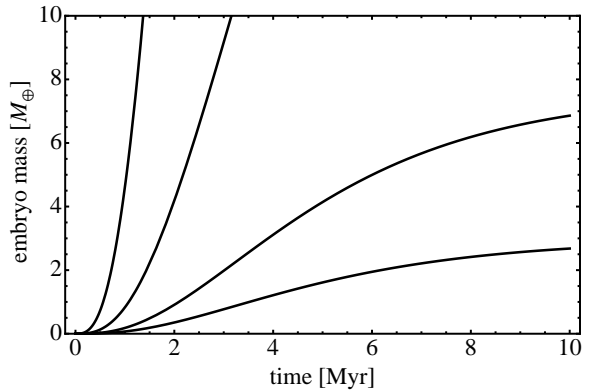}
\caption[growth run2]
	{
	\label{fig:growth2}
	Growth of giant planet embryos in LT, where the envelope temperature, $T_{\rm env}$, has been reduced to $100$ K. The masses of planet embryos, plotted from top to bottom, are Jupiter, Saturn, Neptune and Uranus.
	}

\end{center}
\end{figure}

Figure \ref{fig:growth2} shows the growth of planetary embryos in simulation LT; a disk with a heated envelope temperature of $100$ K. Because of a smaller mass loss rate at the outer boundary, the evolutionary timescale of LT is a factor of $3.6$ over the evolutionary timescale for the disk in the reference model. In this model the cores of Jupiter and Saturn were both able to grow cores of $10$ \Mearth or more during the first $10$ Myr of evolution. This is not surprising considering the prolonged temporal evolution of the disk in the low-temperature model. Despite the success at the growth of the cores of two innermost giant planets, the cores of Neptune and Uranus are unable to grow large enough during the duration of the simulation.

\begin{figure}[ht]
\begin{center}
\includegraphics[width=3in]{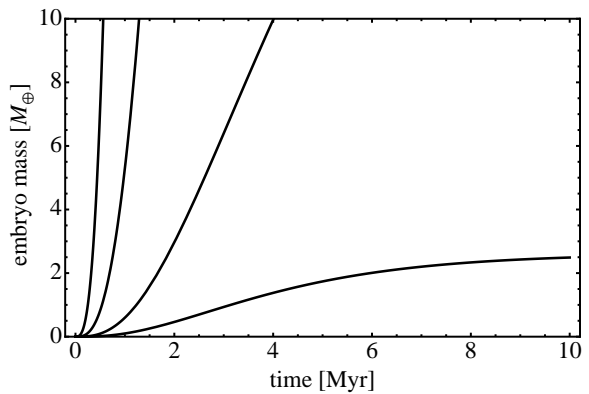}
\caption[growth run4]
	{
	\label{fig:growth4}
	Growth of giant planet embryos in LV, where the viscosity parameter, $\alpha$, has been reduced $0.0001$. The masses of planet embryos, plotted from top to bottom, are Jupiter, Saturn, Neptune and Uranus.
	}
\end{center}
\end{figure}

Figure \ref{fig:growth4} shows the growth of planetary embryos for LV ($\alpha=0.0001$). The embryos in this model grow faster than the embryos in the reference model, and similarly to LT.  All embryos, with the exception of that of the outermost giant planet, are able to grow to  masses of $10$ \Mearth within the allocated $\sim 5-10$ Myr. As seen before in the low-temperature model, the prolonged temporal evolution of the low-viscosity model provided a sufficiently high surface mass density for a long enough time for the three innermost cores to grow to sufficient masses. The effect of varying the viscosity has nearly the same effect on the evolutionary timescale as in the above case where $T_{\rm env}$ was varied, but the truncated disk in LV provides more mass in the giant planet forming region. The extra mass provides a more conducive environment for embryo formation as seen with the success in growing the core of Neptune to $>10$ \Mearth.

As stated earlier, the surface density of solids, $\Sigma_{\rm p}$, was assumed to be $0.014$ times the surface mass density of gas. This estimate is based on the canonical gas/solid ratio of 70 derived from composition of Comet Halley \citep{jessberger88}. It should be noted that this estimate is based solely on the content of ${\rm H_2O}$ ice. Observations of the ejecta of Comet 9P/Tempel 1 during Deep Impact showed significant amounts of CO, ${\rm CO_2}$ and ${\rm CH_3OH}$ \citep{ahearn08}. These ices would certainly be present at the locations of Neptune and Uranus and would result in a higher solid/gas ratio. Combined models of viscous disk evolution and kinetic ice formation show an increase in the solid/gas ratio with radius. By following a chemical reaction network tracing the formation and freeze out of ices in a viscously evolving disk, it was found that the solid surface density at Saturn ($9.5$ AU) is roughly three times that used in previous models of planet formation and that the solid surface density at Uranus ($20$ AU) was higher by a factor of nearly $4.5$ \citep{dodson09}. An increase in solid surface density can also facilitate the formation of planets in a more dramatic fashion. The increase in solids would certainly decrease the formation timescale of the outermost giant planets. Settling of dust to the disk midplane and preferential photoevaporation of gas can lead to a significant increase in the dust-to-gas surface density ratio. This increase in solid surface density can potentially become unstable to gravitational collapse and trigger rapid planetesimal formation \citep{throop05}.


\section{Migration Timescales}
\label{sec:migration}

Viscous torques from density waves launched in a disk from an orbiting planet are thought to cause migration \citep{ward97,ward98}. Using numerical results, \citet{tanaka02} were able to constrain analytical models for the torque exerted by corotational and Lindblad resonances on a body orbiting in an isothermal disk. The net torque in three-dimensions is
\begin{equation}
\label{eqn:torque}
\Gamma = (1.364 + 0.541\frac{d \Sigma_{\rm e}}{d a_{\rm e}})\biggl(\frac{M_{\rm e}}{\Msun}\frac{a_{\rm e} \Omega_{\rm e}}{c_{\rm s}}\biggr)^2 \Sigma_{\rm e} a_{\rm e}^4 \Omega_{\rm e}^2,
\end{equation}
where the subscript ``${\rm e}$'' indicates the values of these variable at location of the embryo. Here, $a_{\rm e}$ refers to the embryo's semimajor axis and $c_{\rm s}$ is the local sound speed of the disk. The local orbital velocity $\Omega_{\rm e}$ is approximated by the Keplerian orbital velocity $\Omega_{\rm kep}$. Eqn. (\ref{eqn:torque}) can then be used to determine the type I migration timescale using
\begin{equation}
\label{eqn:migration}
t_{\rm mig} = \frac{L_{\rm e}}{2 \Gamma} = \frac{M_{\rm e} (G \Msun R_{\rm e})^{1/2}}{2 \Gamma}.
\end{equation}

\begin{figure}[ht]
\begin{center}
\includegraphics[width=3in]{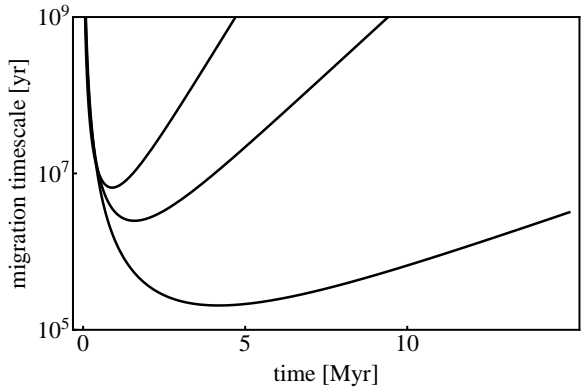}
\caption[migration_temp]
	{
	\label{fig:migrate_temp}
	Jupiter's migration timescales for our reference model and models where the envelope temperature has been varied. From top to bottom, the migration rates shown are from HT, the reference model and LT.
	}

\end{center}
\end{figure}

Figure \ref{fig:migrate_temp} shows Jupiter's migration timescales for LT and HT, models where the envelope temperature, $T_{\rm env}$, has been varied along with our reference model for comparison. They have been calculated keeping the semimajor axis of Jupiter fixed at $5.45$ AU. In general, the simulations that grew planets the fastest also suffered the shortest migration timescales. At early times, the migration timescale decreases due to the growth of the planet, but at late times the decaying surface density of the disk causes the migration timescale to increase. In our reference model, it is unlikely that Jupiter would survive orbital decay due to type I migration. However, if it can survive through the period in which type I migration timescales reach a minimum it has a chance for long term survival. 

\begin{figure}[ht]
\begin{center}
\includegraphics[width=3in]{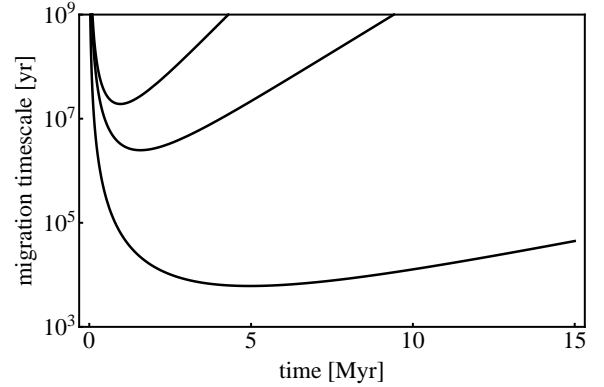}
\caption[migration_visc]
	{
	\label{fig:migrate_visc}
	Jupiter's migration timescales for our reference model and models where the viscosity parameter $\alpha$ has been varied. From top to bottom, the migration rates shown are from HV, the reference model and LV.
	}
\end{center}
\end{figure}

Figure \ref{fig:migrate_temp} shows that the migration timescales of Jupiter for the reference model are intermediate to those in LT and HT. The migration rates in these simulations are mainly affected by the large timescale variations in the evolution of the surface density of the disk. The long (short) timescales produced by lowering (raising) the disk envelope temperature allow planets to grow faster (slower) and maintain surface densities at higher (lower) levels. The combined effect of larger (smaller) planets and higher (lower) surface densities combines to cause shorter (longer) migration timescales than in our reference model. The migration rates of LV and HV are similar to those in LT and HT. Figure \ref{fig:migrate_visc} again shows that models with short evolutionary timescales produce smaller embryos in a less massive disk and therefore these planetary cores have longer migration timescales. 

The migration rates derived using Eqn. (\ref{eqn:migration}) are for type I migration and are only valid when the Hill radius of an embryo is smaller than the scale height of the disk.  When the planet mass exceeds some critical value, the migration switches to type II migration and the evolution of the embryo's semimajor axis becomes locked into the viscous evolution of the disk. Our disk, with its large amount of outward transport of material and truncated outer radius, could cause an embryo to migrate either inward or outward depending on the semimajor axis of a given embryo \citep{ward03}. 

\begin{figure}[ht]
\begin{center}
\includegraphics[width=2.5in]{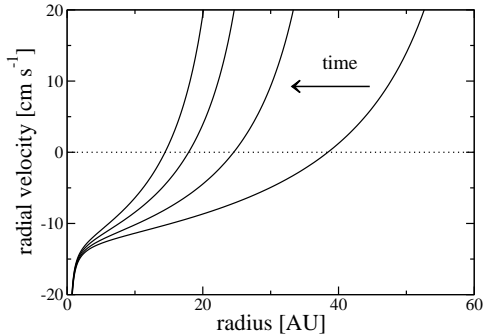}
\caption[velocity]
	{
	\label{fig:velocity}
The radial velocity of material in the disk in the reference model. The four times plotted in Figure \ref{fig:sigma_R1} have also been plotted. They are sequential in time from right to left. The critical radius lies at $38$ AU, $25$ AU, $18$ AU, and $14$ AU at the four times shown. The times shown occur at $0$ Myr, $0.70$ Myr, $1.1$ Myr and $1.3$ Myr respectively.  
	}
\end{center}
\end{figure}

In a truncated disk with outward mass transport, there is a critical radius inward of which the mass moves inward and is accreted onto the central object and outward of which the material is transported outward and eventually out of the system. The survival of a growing embryo against the effects of type II migration depends on which side of the critical radius it is. To investigate the location of the critical radius in our model we have calculated the radial velocity of the material in our disk: 
\begin{equation}
\label{eqn:radial_vel}
v_{\rm r} = -\frac{3}{\Sigma\ r^{1/2}} \frac{\partial}{\partial r}[\nu \Sigma\ r^{1/2}],
\end{equation}
which in terms of our non-dimensional variables is
\begin{equation}
\label{eqn:radial_vel_nondim}
v_{\rm r} = -\frac{3 \nu_0}{R_0}\biggl[\frac{\tilde{h}}{2 \tilde{g}} \frac{\partial \tilde{g}}{\partial \tilde{h}} + \frac{3}{2}\biggr].
\end{equation}
The results of these calculations for the reference model can be seen in Figure \ref{fig:velocity}. We have plotted the radial velocity of the flow for the four times shown in Figure \ref{fig:sigma_R1}. The critical radius lies at $38$ AU, $25$ AU, $18$ AU, and $14$ AU at the four times shown. The times shown occur at $0$ Myr, $0.70$ Myr, $1.1$ Myr, and $1.3$ Myr, respectively. The critical radius, where the curves intersect $v_{\rm r}=0$, moves inward with time. The critical radius at the times shown is exterior to the orbit of Jupiter, but it is rapidly moving inward and will at some point transition to a radius smaller than Jupiter's semimajor axis.  If this happens early enough it could save Jupiter from migrating into the Sun. This could also affect the radial diffusion of dust particles due to turbulent fluctuations in the disk.

\section{Discussion}
\label{sec:discussion}

The five simulations presented here were designed to have comparable masses throughout the planet-forming region ($2\ {\rm AU} < r < 30\ {\rm AU}$). Although their radial surface density profiles are nearly identical within the giant planet-forming region, their profiles are very different in the outer regions of the disk. The various disk models have very different radial profiles at large radii, but one must keep in mind that there is in actuality very little mass in these regions. Despite the very low mass surface densities at large radii, these disks can extend, in some cases, many hundred of AU from the Sun and may have strong implications for the development and evolution of the Kuiper Belt and trans-Neptunian objects \citep{adams04}. 


Despite many of our simulations having large outer disk radii, two simulations, HT and LV have outer disk radii of roughly $20-30$ AU. This is within the current semi-major axis of Neptune. The simulation HT has a disk evolution timescale that is much shorter than our reference model and fails to grow the giant planets within the timescale of gas dissipation. In contrast, LV had a longer disk evolution timescale than our reference model and was more effective at growing the giant planet embryos. The combination of low viscosity and relatively high FUV flux prevented the gas from spreading outward and created a steeper more compact disk. These results show that photoevaporation can affect the solar nebula in the giant planet formation region. 

We have run our models assuming a constant external FUV flux. There are compelling reasons to believe that the incident flux is not constant. Individual stars in the solar birth cluster would certainly have experienced motion relative to one another. Relative motion between the Sun and any of its high-mass brethren would certainly have caused the FUV flux to vary \citep{proszkow09,adams10}. The observed spread in stellar ages within young clusters is usually $\sim1$ Myr, but subgroups within the same cluster have been observed to have a roughly $10$ Myr difference in ages \citep{jeffries06}. This would imply a highly varying UV environment as new stars are born into clusters with the early-type stars rapidly moving onto the main sequence on timescales of $10^4-10^5$ yr. 

The lifetimes of early-type stars are comparable to the lifetimes of circumstellar disks and therefore any changes in luminosity experienced during their short lives will affect the local UV environment. The FUV flux from early-type stars can vary substantially on very short timescales ($10^4\ {\rm yr}$) as they transition through the luminous blue variable (LBV) stage. LBV stars undergo a phase marked by high mass loss and instability. During the LBV phase, a constant bolometric luminosity ($L \approx  10^5 - 10^6\ L_{\odot}$) is maintained, but the dense stellar wind absorbs EUV such that the majority of the flux escapes in the FUV. Observations of proplyds in the Carina Nebula (NGC 3372) suggest these outbursts can have dramatic consequences for nearby protoplanetary disks \citep{smith03}. This would imply direct consequences on disk evolution and survival. In the future, it would be interesting to model a variety of FUV irradiation scenarios and investigate the effect of variable flux rates.

In our models, $\alpha$ was held constant throughout the entire disk and $\nu \propto r$ which follows from $T \propto r^{-1/2}$. In reality, $\nu$ should not be a simple monotonic function if the ionization fraction of the disk varies with radius leading to dead zones where the magnetorotational instability is suppressed. We plan, in the future, to  model viscosity in a more self-consistent manner by utilizing a disk model with an evolving radial temperature profile. This would at first involve using the $\alpha$-viscosity prescription with constant $\alpha$, but with viscosity $\nu$ that is a function of the local disk temperature. By allowing the viscosity to vary throughout the disk it could substantially alter the rates of planet formation in various regions \citep{kretke09,lyra09}. Without further investigation, it is hard to say exactly what the effect of allowing the viscosity to depend on the local qualities of the disk would have on the growth rates and chances for survival against migration of growing embryos. 

Furthermore, our use of a constant $\alpha$ could explain the discrepancy found between the slope of the radial surface density profile of our models and that derived by \citet{desch07}. If $\alpha$ was allowed to be lower in the inner portion of the disk than in the outer regions, then the shallow profile that our models have produced would probably not develop. In this case, it is possible that the steep profile derived by \citet{desch07} would develop. At this time, our numerical model is unable to handle a change of $\alpha$
in the inner portion of the disk and accommodating such a change would require significant modification to our code. We think this is an interesting proposition and would like to investigate it further in the future. Hopefully, we can facilitate such a change when we modify the code to allow for the viscosity to depend on the local disk parameters as stated above.

At any heliocentric distance, the timescale for the growth of dust grains into planetesimals is a few thousand times the local orbital period \citep{weidenschilling00}. Therefore, planetesimals formed in inner regions of a disk form faster than planetesimals in outer regions. Coincidentally, in the inner regions of the planet-forming zone our disk in the reference model matches with the surface density constraints from \citet{desch07} at early times and in the outer planet-forming regions at late times. Any planetesimals formed early on in the inner region, large enough to decouple from the disk yet not so large that they undergo significant migration, will be present and available for planet formation at later times. These planetesimals would effectively maintain the surface density of solids at the high levels needed to match Desch's (2007) constraints after the gas has been transported elsewhere. Farther out in the disk, where planetesimals form more slowly and at later times, our surface density in this region is consistent with surface density constraints also  at later times. Although we have begun the growth of all of the giant planets in our simulations at the same time in each model, there is no reason for this to be the case. It could very well be that planet formation was delayed in one region or another.

In the simulations LV and LT, we are able to successfully grow the embryos of the outer most giant planets within the given $\sim 5-10$ Myr time constraint. It is important to again stress the uncertainties involved in planetary growth from the accumulation of planetesimals. At some point, a transition to oligarchic growth and a clearing of the embryos' feeding zones would slow embryo growth. On the other hand, other processes exist which would increase embryo growth rates. We model growth without atmospheres, from a single size distribution. We also neglect any tidal effects that could dissipate orbital energy from the impacting planetesimals. These are but a few of the major uncertainties in core accretion models. Any of these and others could easily alter the growth rates by a factor of 2 \citep{desch07}. 

We have also made the simplifying assumption that the solids-to-gas ratio is constant with radius. The solids-to-gas ratio should increase with radius as exotic ices condense in the cold outer regions of the solar nebula. The increase of solids beyond the snow line would provide more material for planet growth and decrease the growth timescale of the giant planets, especially of Neptune and Uranus. Photoevaporation of hydrogen and helium from the disk would also tend to increase the solids-to-gas ratio of the outer disk. Despite these uncertainties, we have shown that given reasonable model parameters the cores of the outermost giant planets can successfully be built in a {\it decreting} disk with a truncated outer boundary. It is the outward flow of mass and removal at some outer radius that provides sufficient mass to the outer regions of the disk for planetary core growth.

We placed our growing embryos in the ad hoc compact configuration of the Nice model, but one must keep in mind that the Nice model begins after gas has dissipated. Planet disk interactions would likely have lead to some amount of radial migration while gas was still present in sufficient quantities. This would imply that the giant planets could have begun to grow elsewhere in the disk and migrated to a more compact configuration before the gas dissipated. Our placement of the giant planets at the locations of the Nice model is most likely incorrect, but it is a good proxy for testing planet formation in a compact configuration. A more realistic treatment would require coupling an $N$-body code to our viscous evolution code and evolving it with migration such that the giant planets end up in the compact configuration of the Nice model. We plan to investigate this avenue of study in the near future.

It was our aim to simply illustrate how the migration timescales are affected by the decreasing surface density. One must also keep in mind that type I migration is a poorly understood phenomenon. Most of the studies to date have investigated type I migration in isothermal disks.  Accurately coupling an $N$-body code to our viscous disk model with migration will be difficult enough without the inherent uncertainties in migration processes themselves. It has been shown that in non-isothermal disks with high opacities the induced net torque may have opposite sign and act to push planets outward \citep{paardekooper06}. That is to say, it is uncertain in which direction type I migration would force a planet. Furthermore, recent simulations suggest that the inward scattering of planetesimals could drive the outward migration of growing embryos \citep{levison10}. Type II migration is better understood. It is likely that the largest giant planet Jupiter would quickly fall into the regime of type II migration and be swept inward while on the inside of the critical radius, where the radial gas flow transitions from inward to outward. If, however, the planetary embryo does not completely open a gap the type II migration rate can be reduced or even reversed \citep{crida07}. In our model, the critical radius is continually moving inward. In our reference model, it moves from $38$ AU to $14$ AU over the $1.3$ Myr of disk evolution. At some point, the critical radius should overtake Jupiter and reverse the course of its migration outward. It may be that the decreasing surface density and outward mass transport could save Jupiter from being lost into the Sun.


\section{Summary}
\label{sec:summary}

Photoevaporation has for some time now been invoked as a mechanism for the rapid dispersal of protoplanetary disks \citep{bally98, johnstone98, clarke01, adams04}. It has recently been invoked as a possible mechanism for the truncation of the solar nebula in such a fashion as to produce the steep surface density profile required to produce the giant planets in the compact configuration of the Nice model \citep{desch07}. We have performed a number of simulations to test the relative importance of external photoevaporation versus viscous evolution. We also investigate whether or not photoevaporative mass loss from the outer edge of an evolving protoplanetary disk can produce the steep surface density profile posited by \citet{desch07}. We find, for reasonable disk parameters, that the viscous evolution and photoevaporation play equally important roles in determining disk evolution timescales and morphology.

We have adapted a method of solving propagating phase change problems and developed a one-dimensional viscous disk model that self-consistently tracks the location of the outer boundary under the influence of photoevaporation from an external source. Our application of the formalism of the Stefan problem to astrophysical disks is a novel approach. Our model is, as far as we know, the first model to track the location of the outer boundary in a fully self-consistent manner. We use radial surface density outputs from our viscous disk model to model the growth of giant planet embryos in the compact configuration set forth in the Nice model. Our model is a further exploration of the steady state {\it decretion} disk proposed by \citet{desch07}. With the exception of LV, all of our surface density profiles are shallower than the surface density profile of Eqn. (\ref{eqn:desch}), with a typical dependence on radius of $\Sigma(r) \propto r^{-1.25^{+ 0.88}_{-0.33}}$. Despite the mismatch, our profiles seem conducive to planet formation and do match the surface density constraints derived by \citet{desch07} at certain radii at certain times.

We present five simulations designed to test the effects that varying the strength of the FUV flux and altering the strength of the viscosity have on the temporal evolution of the disk. It took the disk in LT $3.6$ times as long to evolve from a mass of $0.07\ \Msun$ to a mass of $0.0175\ \Msun$ as the disk in our reference model, whereas the disk evolution in HT was faster than of our reference model by a factor of $0.45$. The evolutionary timescale of LV was $3.6$ times longer than the timescale of our reference model and the timescale of HV was $0.81$ times shorter than that of our reference model. The embryos in LT and LV grow faster than the embryos in our reference model, and in LV the cores of the three innermost giant planets were able to grow to $>10$ \Mearth within the allocated $\sim 5-10$ Myr.

The strength of the viscosity and the amount of FUV flux (envelope temperature) were both able to affect the evolutionary timescales of the disks produced in various simulations. A small FUV flux or low viscosity were both found to produce longer evolutionary timescales than our reference model. Both models were more successful than our reference model in growing the giant planet cores, but they produced very different radial surface density profiles. The low FUV flux model produced a radially extended disk whereas the low viscosity model produced a radially contracted disk with a higher surface density in the giant planet forming region. The differences in these two models may provide a natural explanation for the location of the outer edge of the solar system.

The disks produced with our numerical model are all characterized by outward mass transport, mass loss at the outer edge, and a truncated outer boundary. The outer boundary is characterized by substantial mass loss due to photoevaporative heating. This mass loss drives outward mass flow from the critical radius to the outer edge of the disk. The transport of mass from small radii to large can potentially prevent the rapid inward migration of Jupiter and Saturn, while at the same time supply enough mass to the outer regions of the disk for the formation of Uranus and Neptune.


\acknowledgements

The authors thank John Bally for his illuminating discussions and insightful comments. We thank the anonymous referee for helpful comments that improved the quality of the paper. This work was supported by the Cassini Project.

\appendix

\section{Variable Space Grid Method}
\label{subsec:vsg} 

A well-posed problem in the material sciences, called the Stefan Problem, deals with propagating phase changes typically considered in the context of melting/freezing and heat ablation \citep{ozisik80}. We have developed a 1-D model of a protoplanetary disk that includes viscous diffusion and photoevaporation at the outer boundary using numerical techniques developed to solve the Stefan Problem. By adapting the Stefan problem to astrophysical disks, our 1-D numerical model self-consistently tracks the location of the outer boundary. This is a novel approach to modeling the solar nebula.

To solve the Stefan problem, \citet{kutluay97} adopt a numerical method with a variable space grid (VSG) first proposed by \citet{murray59}. The VSG method employs a fixed number of grid points with a variable grid size at each time step. This method involves solving two coupled differential equations at each time step, one for the location of the outer boundary and one for the diffusive evolution of the disk. Once the location of the outer boundary is found the abscissa is rescaled and the diffusive evolution calculated.

We will now briefly describe the VSG method. For a non-dimensional diffusion equation,
\begin{equation}
\label{eqn:heat}
\frac{\partial U}{\partial t} = \frac{\partial^2 U}{\partial x^2}
\end{equation}
with the boundary condition
\begin{equation}
\label{eqn:boundary}
\frac{ds(t)}{dt} = f(x,U_x)
\end{equation}
where $s(t)$ is the location of the boundary at time $t$. Eqn. (\ref{eqn:heat}) can be differentiated with respect to time and, for the $i$th grid point,
\begin{equation}
\label{eqn:ithgrid}
\frac{\partial U}{\partial t}\Biggr|_i=\frac{\partial U}{\partial x}\Biggr|_t\frac{\partial x}{\partial t}\Biggr|_i+\frac{\partial{U}}{\partial t}\Biggr|_x
\end{equation}
assuming the grid point $x_i$ is moved by
\begin{equation}
\label{eqn:node}
\frac{dx_i}{dt}=\frac{x_i}{s(t)}\cdot \frac{ds}{dt}.
\end{equation}

By substituting Eqn. (\ref{eqn:node}) into Eqn. (\ref{eqn:ithgrid}), the heat equation (Eqn. (\ref{eqn:heat})) can be reformulated as
\begin{equation}
\label{eqn:kutluay}
\frac{\partial U}{\partial t} = \frac{x_i}{s(t)} \frac{ds(t)}{dt} \frac{\partial U}{\partial x} + \frac{\partial^2 U}{\partial x^2}.
\end{equation}

Various methods, such as similarity solutions, have been used to solve Stefan problems analytically. We used these solutions to test our numerical code. One such test was done on the following system, a Stefan problem of transient heat conduction in a melting slab \citep{bluman74}. The one-dimensional, finite slab is insulated at $x=0$ and has a propagating phase change at $x=X(t)$, where heat flows into the melting face at a rate $H(t)$.
\begin{equation}
\label{eqn:self1}
\frac{\partial u}{\partial t} = \frac{\partial^2 u}{\partial x^2},\ 0<x<X(t)
\end{equation}
with the boundary condition
\begin{equation}
\label{eqn:self2}
H(t) = \frac{\partial u}{\partial x} - \frac{dX}{dt}.
\end{equation}

Given the following boundary and initial conditions
\begin{equation}
\label{eqn:self3}
u(0,t) = 0,\ x=0
\end{equation}

\begin{equation}
\label{eqn:self4}
u(X(t),t) = 0,\ x=X(t)
\end{equation}
and
\begin{equation}
\label{eqn:self5}
u(x,0) = g(x)
\end{equation}
and assuming $X(t) = 1-t$, the exact solution is of a self-similar form 
\begin{equation}
\label{eqn:self6}
u(x,t) = {\rm exp}(\pi^2)\frac{{\rm sin}(\pi \xi)}{(1-t)^{1/2}}{\rm exp}\biggl(\frac{-\pi^2}{1-t}-\frac{\xi^2 (t-1)}{4}\biggr)
\end{equation}
where $\xi = x/X(t)$.

The exact analytic, self-similar solution was used to check the numerical VSG method. We tested our code for convergence against the exact analytical solutions by increasing the spatial grid resolution. These tests were all done with the same sized time steps. The results of these tests have been tabulated in Table \ref{tab:converge}. The numerical results are in good agreement with the exact solution and exhibits the expected convergence as the number of grid point, $N$, increases.

\begin{table}[ht]
\caption[slopes]
	{
	\label{tab:converge}
	\label{firsttab}
Analytic test for the VSG method. Results of the tests for convergence against exact, analytic self-similar solution at a final time of $t=0.25$ are shown.}
\begin{center}
\item[]\begin{tabular}{ccccc}
\tableline 
\tableline
    $x/X(t)$  & $u$ (actual) & $u\ (N=50)$ & $u\ (N=100)$ & $u\ (N=200)$ \\
\hline
    $0.0$ & $0.00000000$ & $0.00000000$ &  $0.00000000$ &  $0.00000000$ \\
    $0.2$ & $0.02547849$ & $0.02540785$ &  $0.02546055$ &  $0.02547405$ \\
    $0.4$ & $0.04216313$ & $0.04204546$ &  $0.04213325$ &  $0.04215575$ \\
    $0.6$ & $0.04377427$ & $0.04365066$ &  $0.04374288$ &  $0.04376651$ \\
    $0.8$ & $0.02851226$ & $0.02843037$ &  $0.02849147$ &  $0.02850713$ \\
    $1.0$ & $0.00000000$ & $0.00000000$ &  $0.00000000$ &  $0.00000000$ \\
\tableline
\end{tabular}
\end{center}
\end{table}

\begin{figure}[ht]
\begin{center}
\includegraphics[width=2.5in]{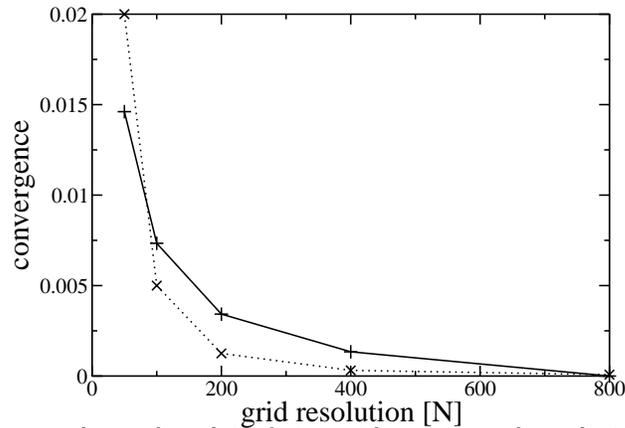}
\caption[sigma_converge]
	{
	\label{fig:sigma_converge}
The spatial convergence of a number of simulations with various numbers of grid spacings. The calculated convergence is shown with plus symbols connected by solid lines. The expected $1/N^2$ convergence is shown with x's connected by dotted lines. In all of our simulations we use $N=200$ grid spaces.
	}
\end{center}
\end{figure}

As a further test of convergence on the problem at hand we have completed a number of simulations with a variety of grid sizes. We have checked for both spatial and temporal convergence. Figure \ref{fig:sigma_converge} shows the spatial convergence as the number of grid spaces increases. The convergence has been calculated by differencing the surface density of each of the lower resolution simulations from the surface density of the highest resolution simulation ($N=800$) and then normalizing by the innermost available grid space. The computed convergence is shown with plus symbols connected by solid lines. For comparison, the expected $1/N^2$ convergence is shown over-plotted with x's connected by dotted lines. It can clearly be seen that the actual convergence is very close to the expected convergence. We have chosen to use $N=200$ for our number of grid spaces. This allows for simulations that complete in a reasonable amount of time and is acceptably accurate, to within less than $0.5 \%$ of the highest resolution simulation. Due to the large uncertainties in many model parameters, we feel this level of convergence is acceptable.

\begin{figure}[ht]
\begin{center}
\includegraphics[width=2.5in]{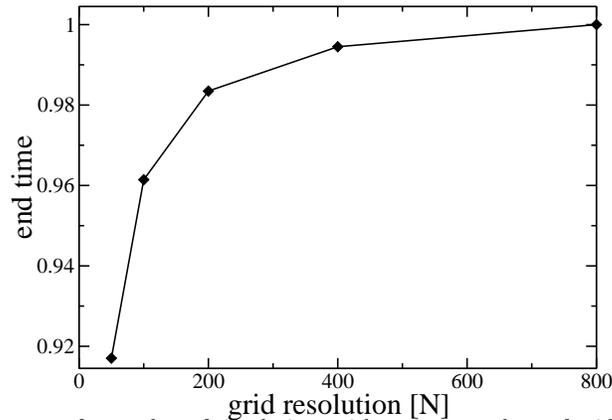}
\caption[time_converge]
	{
	\label{fig:time_converge}
The temporal convergence of a number of simulations with various numbers of grid spacings. In all of our simulations we use $N=200$ grid spaces.
	}
\end{center}
\end{figure}

We have also investigated the temporal convergence of our model. For this test we have divided the end times of each simulation by the end time of the highest resolution simulation. Again, the highest resolution simulation has $N=800$. The temporal convergence was calculated by dividing the final time of each lower resolution simulation by the final time of the highest resolution simulation. It can be seen in Figure \ref{fig:time_converge} that the final time of the simulation with $N=200$ is within $2 \%$ of the final time of the highest resolution simulation. As with the spatial convergence, we feel that this is sufficient considering the large uncertainties in many of our model parameters.

\bibliography{bibliography.bib}

\begin{thebibliography}{64}
\expandafter\ifx\csname natexlab\endcsname\relax\def\natexlab#1{#1}\fi

\bibitem[{{Adams}(2010)}]{adams10}
{Adams}, F.~C. 2010, ArXiv e-prints

\bibitem[{{Adams} {et~al.}(2004){Adams}, {Hollenbach}, {Laughlin}, \&
  {Gorti}}]{adams04}
{Adams}, F.~C., {Hollenbach}, D., {Laughlin}, G., \& {Gorti}, U. 2004, Apj,
  611, 360

\bibitem[{{A'Hearn}(2008)}]{ahearn08}
{A'Hearn}, M.~F. 2008, Space Science Reviews, 138, 237

\bibitem[{{Armitage}(2010)}]{armitage10}
{Armitage}, P.~J. 2010, {Astrophysics of Planet Formation}

\bibitem[{{Bally} {et~al.}(1998){Bally}, {Sutherland}, {Devine}, \&
  {Johnstone}}]{bally98}
{Bally}, J., {Sutherland}, R.~S., {Devine}, D., \& {Johnstone}, D. 1998, AJ,
  116, 293

\bibitem[{{Bluman} \& {Cole}(1974)}]{bluman74}
{Bluman}, G.~W., \& {Cole}, J.~D. 1974, {Similarity methods for differential
  equations}

\bibitem[{{Brasser} {et~al.}(2006){Brasser}, {Duncan}, \&
  {Levison}}]{brasser06}
{Brasser}, R., {Duncan}, M.~J., \& {Levison}, H.~F. 2006, Icarus, 184, 59

\bibitem[{{Chiang} \& {Goldreich}(1997)}]{chiang97}
{Chiang}, E.~I., \& {Goldreich}, P. 1997, ApJ, 490, 368

\bibitem[{{Clarke}(2007)}]{clarke07}
{Clarke}, C.~J. 2007, MNRAS, 376, 1350

\bibitem[{{Clarke} {et~al.}(2001){Clarke}, {Gendrin}, \&
  {Sotomayor}}]{clarke01}
{Clarke}, C.~J., {Gendrin}, A., \& {Sotomayor}, M. 2001, \mnras, 328, 485

\bibitem[{{Crida}(2009)}]{crida09}
{Crida}, A. 2009, ApJ, 698, 606

\bibitem[{{Crida} \& {Morbidelli}(2007)}]{crida07}
{Crida}, A., \& {Morbidelli}, A. 2007, MNRAS, 377, 1324

\bibitem[{{Desch}(2007)}]{desch07}
{Desch}, S.~J. 2007, ApJ, 671, 878

\bibitem[{{Dodson-Robinson} {et~al.}(2009){Dodson-Robinson}, {Willacy},
  {Bodenheimer}, {Turner}, \& {Beichman}}]{dodson09}
{Dodson-Robinson}, S.~E., {Willacy}, K., {Bodenheimer}, P., {Turner}, N.~J., \&
  {Beichman}, C.~A. 2009, Icarus, 200, 672

\bibitem[{{Fatuzzo} \& {Adams}(2008)}]{fatuzzo08}
{Fatuzzo}, M., \& {Adams}, F.~C. 2008, \apj, 675, 1361

\bibitem[{{Gautier} {et~al.}(2001){Gautier}, {Hersant}, {Mousis}, \&
  {Lunine}}]{gautier01}
{Gautier}, D., {Hersant}, F., {Mousis}, O., \& {Lunine}, J.~I. 2001, ApJ, 550,
  L227

\bibitem[{{Gomes} {et~al.}(2005){Gomes}, {Levison}, {Tsiganis}, \&
  {Morbidelli}}]{gomes05}
{Gomes}, R., {Levison}, H.~F., {Tsiganis}, K., \& {Morbidelli}, A. 2005,
  Nature, 435, 466

\bibitem[{{Gorti} {et~al.}(2009){Gorti}, {Dullemond}, \&
  {Hollenbach}}]{gorti09}
{Gorti}, U., {Dullemond}, C.~P., \& {Hollenbach}, D. 2009, \apj, 705, 1237

\bibitem[{{Guillot} \& {Hueso}(2006)}]{guillot06}
{Guillot}, T., \& {Hueso}, R. 2006, MNRAS, 367, L47

\bibitem[{{Haisch} {et~al.}(2001){Haisch}, {Lada}, \& {Lada}}]{haisch01}
{Haisch}, Jr., K.~E., {Lada}, E.~A., \& {Lada}, C.~J. 2001, \apjl, 553, L153

\bibitem[{{Hartmann}(1998)}]{hartmann98a}
{Hartmann}, L. 1998, {Accretion Processes in Star Formation}, ed. L.~Hartmann

\bibitem[{{Hartmann} {et~al.}(1998){Hartmann}, {Calvet}, {Gullbring}, \&
  {D'Alessio}}]{hartmann98b}
{Hartmann}, L., {Calvet}, N., {Gullbring}, E., \& {D'Alessio}, P. 1998, ApJ,
  495, 385

\bibitem[{{Hayashi}(1981)}]{hayashi81}
{Hayashi}, C. 1981, Progress of Theoretical Physics Supplement, 70, 35

\bibitem[{{Hern{\'a}ndez} {et~al.}(2007){Hern{\'a}ndez}, {Calvet},
  {Brice{\~n}o}, {Hartmann}, {Vivas}, {Muzerolle}, {Downes}, {Allen}, \&
  {Gutermuth}}]{hernandez07}
{Hern{\'a}ndez}, J., {et~al.} 2007, \apj, 671, 1784

\bibitem[{{Hollenbach} \& {Adams}(2004)}]{hollenbach04}
{Hollenbach}, D., \& {Adams}, F.~C. 2004, in Astronomical Society of the
  Pacific Conference Series, Vol. 323, Star Formation in the Interstellar
  Medium: In Honor of David Hollenbach, ed. {D.~Johnstone, F.~C.~Adams,
  D.~N.~C.~Lin, D.~A.~Neufeeld, \& E.~C.~Ostriker }, 3--18

\bibitem[{{Jeffries} {et~al.}(2006){Jeffries}, {Maxted}, {Oliveira}, \&
  {Naylor}}]{jeffries06}
{Jeffries}, R.~D., {Maxted}, P.~F.~L., {Oliveira}, J.~M., \& {Naylor}, T. 2006,
  \mnras, 371, L6

\bibitem[{{Jessberger} {et~al.}(1988){Jessberger}, {Christoforidis}, \&
  {Kissel}}]{jessberger88}
{Jessberger}, E.~K., {Christoforidis}, A., \& {Kissel}, J. 1988, Nature, 332,
  691

\bibitem[{{Johnstone} {et~al.}(1998){Johnstone}, {Hollenbach}, \&
  {Bally}}]{johnstone98}
{Johnstone}, D., {Hollenbach}, D., \& {Bally}, J. 1998, ApJ, 499, 758

\bibitem[{{Kenyon} \& {Bromley}(2004)}]{kenyon04}
{Kenyon}, S.~J., \& {Bromley}, B.~C. 2004, \nat, 432, 598

\bibitem[{{Kokubo} \& {Ida}(1996)}]{kokubo96}
{Kokubo}, E., \& {Ida}, S. 1996, Icarus, 123, 180

\bibitem[{{Kokubo} \& {Ida}(1998)}]{kokubo98}
---. 1998, Icarus, 131, 171

\bibitem[{{Kokubo} \& {Ida}(2000)}]{kokubo00}
---. 2000, Icarus, 143, 15

\bibitem[{{Kokubo} \& {Ida}(2002)}]{kokubo02}
---. 2002, ApJ, 581, 666

\bibitem[{{Kretke} {et~al.}(2009){Kretke}, {Lin}, {Garaud}, \&
  {Turner}}]{kretke09}
{Kretke}, K.~A., {Lin}, D.~N.~C., {Garaud}, P., \& {Turner}, N.~J. 2009, \apj,
  690, 407

\bibitem[{{Kutluay} {et~al.}(1997){Kutluay}, {Bahadir}, \&
  {\"Ozdes}}]{kutluay97}
{Kutluay}, S., {Bahadir}, A.~R., \& {\"Ozdes}, A. 1997, JCAM, 81, 135

\bibitem[{{Lada} \& {Lada}(2003)}]{lada03}
{Lada}, C.~J., \& {Lada}, E.~A. 2003, ARA\&A, 41, 57

\bibitem[{{Levison} {et~al.}(2010){Levison}, {Thommes}, \&
  {Duncan}}]{levison10}
{Levison}, H.~F., {Thommes}, E., \& {Duncan}, M.~J. 2010, AJ, 139, 1297

\bibitem[{{Liffman}(2003)}]{liffman03}
{Liffman}, K. 2003, Publications of the Astronomical Society of Australia, 20,
  337

\bibitem[{{Lissauer}(1987)}]{lissauer87}
{Lissauer}, J.~J. 1987, Icarus, 69, 249

\bibitem[{{Lissauer}(1993)}]{lissauer93}
---. 1993, \araa, 31, 129

\bibitem[{{Lynden-Bell} \& {Pringle}(1974)}]{lynden-bell74}
{Lynden-Bell}, D., \& {Pringle}, J.~E. 1974, MNRAS, 168, 603

\bibitem[{{Lyra} {et~al.}(2009){Lyra}, {Johansen}, {Zsom}, {Klahr}, \&
  {Piskunov}}]{lyra09}
{Lyra}, W., {Johansen}, A., {Zsom}, A., {Klahr}, H., \& {Piskunov}, N. 2009,
  \aap, 497, 869

\bibitem[{{Morbidelli} \& {Levison}(2004)}]{morbidelli04}
{Morbidelli}, A., \& {Levison}, H.~F. 2004, \aj, 128, 2564

\bibitem[{{Morbidelli} {et~al.}(2005){Morbidelli}, {Levison}, {Tsiganis}, \&
  {Gomes}}]{morbidelli05}
{Morbidelli}, A., {Levison}, H.~F., {Tsiganis}, K., \& {Gomes}, R. 2005,
  Nature, 435, 462

\bibitem[{{Murray} \& {Landis}(1959)}]{murray59}
{Murray}, W.~D., \& {Landis}, F. 1959, Trans. ASME J. Heat Transfer, 81, 106

\bibitem[{{Owen} {et~al.}(1999){Owen}, {Mahaffy}, {Niemann}, {Atreya},
  {Donahue}, {Bar-Nun}, \& {de Pater}}]{owen99}
{Owen}, T., {Mahaffy}, P., {Niemann}, H.~B., {Atreya}, S., {Donahue}, T.,
  {Bar-Nun}, A., \& {de Pater}, I. 1999, Nat, 402, 269

\bibitem[{{\"{O}zi\c{s}ik}(1980)}]{ozisik80}
{\"{O}zi\c{s}ik}, M.~N. 1980, {Heat Conduction}

\bibitem[{{Paardekooper} \& {Mellema}(2006)}]{paardekooper06}
{Paardekooper}, S.-J., \& {Mellema}, G. 2006, A\&A, 459, L17

\bibitem[{{Proszkow} \& {Adams}(2009)}]{proszkow09}
{Proszkow}, E., \& {Adams}, F.~C. 2009, \apjs, 185, 486

\bibitem[{{Rafikov}(2004)}]{rafikov04}
{Rafikov}, R.~R. 2004, AJ, 128, 1348

\bibitem[{{Safronov}(1969)}]{safronov69}
{Safronov}, V. 1969, {Evolution of the protoplanetary cloud and formation of
  the earth and planets.}

\bibitem[{{Shakura} \& {Sunyaev}(1973)}]{shakura73}
{Shakura}, N.~I., \& {Sunyaev}, R.~A. 1973, A\&A, 24, 337

\bibitem[{{Smith} {et~al.}(2003){Smith}, {Bally}, \& {Morse}}]{smith03}
{Smith}, N., {Bally}, J., \& {Morse}, J.~A. 2003, \apjl, 587, L105

\bibitem[{{Tanaka} {et~al.}(2002){Tanaka}, {Takeuchi}, \& {Ward}}]{tanaka02}
{Tanaka}, H., {Takeuchi}, T., \& {Ward}, W.~R. 2002, ApJ, 565, 1257

\bibitem[{{Throop} \& {Bally}(2005)}]{throop05}
{Throop}, H.~B., \& {Bally}, J. 2005, \apjl, 623, L149

\bibitem[{{Tielens}(2005)}]{tielens05}
{Tielens}, A.~G.~G.~M. 2005, {The Physics and Chemistry of the Interstellar
  Medium}, ed. {Tielens, A.~G.~G.~M.}

\bibitem[{{Tsiganis} {et~al.}(2005){Tsiganis}, {Gomes}, {Morbidelli}, \&
  {Levison}}]{tsiganis05}
{Tsiganis}, K., {Gomes}, R., {Morbidelli}, A., \& {Levison}, H.~F. 2005,
  Nature, 435, 459

\bibitem[{{Wadhwa} {et~al.}(2007){Wadhwa}, {Amelin}, {Davis}, {Lugmair},
  {Meyer}, {Gounelle}, \& {Desch}}]{wadhwa07}
{Wadhwa}, M., {Amelin}, Y., {Davis}, A.~M., {Lugmair}, G.~W., {Meyer}, B.,
  {Gounelle}, M., \& {Desch}, S.~J. 2007, Protostars and Planets V, 835

\bibitem[{{Ward}(1997)}]{ward97}
{Ward}, W.~R. 1997, Icarus, 126, 261

\bibitem[{{Ward}(1998)}]{ward98}
{Ward}, W.~R. 1998, in Astronomical Society of the Pacific Conference Series,
  Vol. 148, Origins, ed. {C.~E.~Woodward, J.~M.~Shull, \& H.~A.~Thronson Jr.},
  338--346

\bibitem[{{Ward}(2003)}]{ward03}
{Ward}, W.~R. 2003, in Lunar and Planetary Institute Science Conference
  Abstracts, Vol.~34, Lunar and Planetary Institute Science Conference
  Abstracts, ed. {S.~Mackwell \& E.~Stansbery}, abstract no. 1736

\bibitem[{{Weidenschilling}(1977)}]{weidenschilling77}
{Weidenschilling}, S.~J. 1977, Ap\&SS, 51, 153

\bibitem[{{Weidenschilling}(2000)}]{weidenschilling00}
---. 2000, Space Science Reviews, 92, 295

\bibitem[{{Wetherill} \& {Stewart}(1989)}]{wetherill89}
{Wetherill}, G.~W., \& {Stewart}, G.~R. 1989, Icarus, 77, 330

\end{thebibliography}
\bibliographystyle{apj}

\clearpage

\end{document}